\documentclass[prd,,aps,nofootinbib,amsmath,amssymb,superscriptaddress]{revtex4}
\usepackage{amssymb,amsmath,epsfig,bm}
\usepackage[usenames,dvipsnames]{color}
\usepackage[bookmarks]{hyperref}

\newtheorem{lemma}{Lemma}
\newtheorem{proposition}{Proposition}
\newtheorem{theorem}{Theorem}

\newcommand{\proof}{\noindent {\bf Proof. }}
\newcommand{\proofof}[1]{\noindent {\bf Proof of #1. }}
\newcommand{\qed}{\hfill $\fbox{\hspace{0.3mm}}$ \vspace{.3cm}} 

\newcommand{\Real}{\mathbb{R}}
\newcommand{\Complex}{\mathbb{C}}
\newcommand{\ve}[1]{\underline{#1}}
\newcommand{\re}{\mbox{Re}}
\newcommand{\im}{\mbox{Im}}

\begin{document}

\title{Cauchy horizon stability in a collapsing spherical dust cloud. II: Energy bounds for test fields and odd-parity gravitational perturbations}

\author{N\'estor Ortiz}
\email{nortiz@perimeterinstitute.ca}
\affiliation{Perimeter Institute for Theoretical Physics, 31 Caroline St., Waterloo, ON, N2L 2Y5, Canada.}
\author{Olivier Sarbach}
\email{sarbach@ifm.umich.mx}
\affiliation{Instituto de F\'isica y Matem\'aticas, Universidad Michoacana de San Nicol\'as de Hidalgo, Edificio C-3, Ciudad Universitaria, 58040, Morelia, Michoac\'an, M\'exico.}

\date{\today}

\begin{abstract}
We analyze the stability of the Cauchy horizon associated with a globally naked, shell-focussing singularity arising from the complete gravitational collapse of a spherical dust cloud. In a previous work, we have studied the dynamics of spherical test scalar fields on such a background. In particular, we proved that such fields cannot develop any divergences which propagate along the Cauchy horizon. In the present work, we extend our analysis to the more general case of test fields without symmetries and to linearized gravitational perturbations with odd parity. To this purpose, we first consider test fields possessing a divergence-free stress-energy tensor satisfying the dominant energy condition, and we prove that a suitable energy norm is uniformly bounded in the domain of dependence of the initial slice. In particular, this result implies that free-falling observers co-moving with the dust particles measure a finite energy of the field, even as they cross the Cauchy horizon at points lying arbitrarily close to the central singularity. Next, for the case of Klein-Gordon fields, we derive point-wise bounds from our energy estimates which imply that the scalar field cannot diverge at the Cauchy horizon, except possibly at the central singular point. Finally, we analyze the behaviour of odd-parity, linear gravitational and dust perturbations of the collapsing spacetime. Similarly to the scalar field case, we prove that the relevant gauge-invariant combinations of the metric perturbations stay bounded away from the central singularity, implying that no divergences can propagate in the vacuum region. Our results are in accordance with previous numerical studies and analytic work in the self-similar case.
\end{abstract}

\maketitle

\section{Introduction}
\label{Sec:Intro}

The Tolman-Bondi (TB) spacetime provides an exact, dynamical solution of Einstein's field equations sourced by a spherical, zero pressure perfect fluid. (See, for instance Ref.~\cite{Griffiths-Podolsky-Book} for a summary on the properties of these spacetimes.) Therefore, it constitutes a simple relativistic model for the gravitational collapse of a dust star, and historically it has played an important role in shedding light on the understanding of black hole formation~\cite{jOhS39}. For more than four decades now, it has been known that, for an inhomogeneous initial density distribution, this model leads to the formation of shell-crossing or shell-focusing singularities, a portion of which is causally connected to local observers~\cite{pYhShM73,dElS79,dC84}. Although this collapse model is hardly realistic since angular momentum, pressure gradients and other effects taking place in a real star are neglected, it is still relevant from the theoretical point of view. On the one hand, it has the appealing property that, in co-moving synchronous coordinates, the geometric and fluid quantities can be represented in closed explicit form, which opens the gate to an understanding of its physical features by rigorous means. On the other hand, there exists a non-zero measure class of initial data within this model for which the resulting singularity turns out to be causally connected to future null infinity, and thus {\it globally naked} in the sense that a portion of the null, shell-focusing singularity can be ``seen" by observers moving in the asymptotic region of the spacetime in a finite proper time~\cite{dC84,rN86,pJiD93,Joshi-Book,nOoS11}\footnote{For potentially observable properties of TB globally naked singularities we refer the reader to Refs.~\cite{kNnKhI03,lKdMcB14,nOoS14b,nOoStZ15,nOoStZ15b}. Observational features such as weak gravitational lensing have also been studied for naked singularities with a different structure than the ones studied here, see Refs.~\cite{kVgE02,kVdNsC98,kVcK08}.}. This fact naturally questions the validity of the Weak Cosmic Censorship (WCC) conjecture~\cite{rP69,rW97}. According to the WCC conjecture, globally naked singularities should be unstable under generic, non-spherical perturbations of the initial data or when realistic matter models are considered. However, despite several efforts, this stability problem remains open in the case of TB collapse\footnote{An extended historical review on this stability problem can be found in the introduction of our previous work~\cite{nOoS14}.}. Therefore, the TB model offers a suitable theoretical platform to study the validity of the WCC conjecture in a four-dimensional asymptotically flat spacetime. In an attempt to gain further insight into the stability problem of naked singularities in TB spacetimes, in a previous work~\cite{nOoS14} we considered test fields propagating on the fixed---but dynamical---spacetime given by the TB geometry, and studied the dynamics of such test fields in a vicinity of a central globally naked singularity and its associated Cauchy horizon, which corresponds to the first light ray that emanates from the singularity and extends all the way to future null infinity. As we only consider smooth initial data for the test field's evolution, a divergent behaviour originating at the central singularity and propagating along the Cauchy horizon would suggest an instability of such horizon if back reaction of the field would be taken into account. Our investigations in~\cite{nOoS14} first considered test fields in the geometric optics approximation, and we showed that although they undergo a blueshift along the Cauchy horizon, this blueshift is uniformly bounded. Subsequently, we studied the Cauchy evolution of a spherically symmetric test scalar field, which we proved to be everywhere finite on the Cauchy horizon away from the central singularity.

The goal of the present article is to gain a deeper insight into the stability properties of the Cauchy horizon associated with naked singularities in TB spacetimes. For this purpose, we take a step beyond the analysis of spherically symmetric test fields, and consider more general scenarios, including the propagation of electromagnetic test fields, massive scalar test fields with arbitrary angular momentum, and linearized odd-parity metric perturbations of TB spacetimes. The main difficulty when tackling non-spherical fields is the presence of the centrifugal term (of the typical form $\ell(\ell+1)/r^2$ with $\ell$ the angular momentum number and $r$ the areal radius) in the effective potential governing the field propagation, which is non-integrable in the vicinity of the central singularity, and thus cannot be treated using the main results of our previous work, which was based on the characteristic approach.

Therefore, in this work we change the strategy and address the problem using {\it energy estimates} techniques. {\it Grosso modo}, the prescription is the following. Assume that the field of interest possesses a stress-energy tensor ${\bf T}$ which has zero divergence (as is always the case if the field arises from a fundamental physical theory and satisfies the equations of motion), and assume that ${\bf T}$ satisfies the dominant energy condition. We then specify a future-directed timelike vector field ${\bf X}$ with associated future-directed causal current density ${\bf J}_{\bf X}$ defined by $J_{\bf X}^{\mu} := -T^{\mu}{}_\nu X^{\nu}$. Given a spacelike hypersurface $\Sigma$, the ``energy" ${\cal E}_{\bf X}[\Sigma]$ contained in $\Sigma$, can be defined as minus the flux of ${\bf J}_{\bf X}$ through $\Sigma$ (which is guaranteed to be nonnegative due to the dominant energy condition). Via Gauss' theorem, one obtains inequalities which relate the energies ${\cal E}_{\bf X}[\Sigma_1]$ and ${\cal E}_{\bf X}[\Sigma_2]$ belonging to different Cauchy surfaces $\Sigma_1$ and $\Sigma_2$. The key step is to find good choices for the vector field ${\bf X}$ which yield the best possible inequalities and/or energy expressions ${\cal E}_{\bf X}$. For example, when showing stability one seeks to prove that the energy ${\cal E}_{\bf X}$ stays bounded or decays in time and that it provides a norm that is sufficiently strong to bound the desired features of the solution.

When considering a (globally) stationary spacetime $(M,{\bf g})$, there is a natural choice for the vector field ${\bf X}$ which is given by the Killing vector field generating the time symmetry, in which case ${\bf J}_{\bf X}$ is divergence-free and the energy ${\cal E}_{\bf X}$ is conserved in time. However, there are many interesting situations in general relativity where no such preferred or appropriate choices exist. This is already the case for stationary black holes, where the asymptotically timelike Killing vector field either becomes null at the horizon (Schwarzschild) or becomes spacelike inside the ergosphere (Kerr), and thus does not provide a global timelike Killing vector field. Energy estimates involving modified vector fields ${\bf X}$ are then required to prove stability results controlling the fields and their derivatives, including the ones transverse to the horizon, see for instance~\cite{mDiR08,mDiRyS16,mDgHiR16,lApBjJ16} for recent results. For the dynamical case of TB spacetimes with a globally naked singularity, it is even less clear how to choose a good vector field ${\bf X}$ since there are no timelike Killing vector fields to begin with. 

Surprisingly, it turns out that in spite of these difficulties there does exist a choice for ${\bf X}$ which yields a {\it uniform energy bound} for ${\cal E}_{\bf X}$, even though the spacetime is dynamical and singular in the interior of the cloud. This choice is provided by the {\it Kodama vector field}~\cite{hK80} which can be defined in any spherically symmetric spacetime (static or not) and yields a locally conserved current whose associated conserved quantity turns out to be the Misner-Sharp~\cite{cMdS64} or Hawking~\cite{cMdS64} mass function. In this work, we prove that the use of the Kodama vector field provides a uniform bound on ${\cal E}_{\bf X}$ for the case of TB spacetimes describing a collapsing cloud of finite radius forming a globally naked singularity. In particular, our result implies a uniform bound for the energy of test fields measured by observers co-moving with the dust particles.

Combining our energy bounds with Sobolev-type estimates, we also bound the amplitude of scalar test fields and of linearized gravitational odd-parity perturbations propagating on the collapsing background spacetime. The main result of this work consists of a collection of theorems which establish that (assuming smooth initial data which is compactly supported on a Cauchy surface) such fields cannot grow arbitrarily large at the Cauchy horizon, with the possible exception of the central singular point from which the Cauchy horizon emanates. Therefore, even though our bounds do not exclude the situation in which the field diverges at the central singular point, they do exclude the possibility that such a hypothetical divergence could propagate along the Cauchy horizon to the exterior region.
In fact, our bounds show that the field must decay along the Cauchy horizon. Our rigorous results are in qualitative agreement with previous numerical work in~\cite{hItHkN98,hItHkN99} which found that in the marginally bound case of spherical dust collapse, a globally naked singularity cannot act as a strong source of odd-parity gravitational radiation, even though the metric perturbations grow in the central region. Our results are also in agreement with previous analytic work regarding the stability of the Cauchy horizon with respect to odd-parity linear perturbations of self-similar TB collapse, which admits an additional homothetic vector field~\cite{eDbN11b}.

The remaining of this article is organized as follows. In the next section, we review the basic properties of the TB collapse model, state our assumptions on the initial data, and establish some preliminary results on the behaviour of relevant fields close to the singularity. In section~\ref{Sec:EnergyEstimates} we introduce the energy norms ${\cal E}_{\bf X}$ corresponding to a given future-directed timelike vector field ${\bf X}$, discuss energy balance laws, and next we state and prove our main theorem concerning the uniform energy bound. In section~\ref{Sec:Applications} we present three applications. In subsections~\ref{Sub:Electromagnetic_fields}~and~\ref{Sub:Scalar_fields} we provide examples corresponding to electromagnetic and linear scalar fields, respectively. Last, in section~\ref{Sub:GravPert}, we consider the case of odd-parity, linear gravitational and dust perturbations. We summarize our results and discuss their significance for the stability of globally naked singularities arising from the spherically symmetric dust collapse model in section~\ref{Sec:Discussion}. The proofs of some technical statements used in this work are given in appendices.

Throughout this work we use the signature convention $(-,+,+,+)$ for the metric, and we use geometrized units in which the gravitational constant and speed of light are both set equal to one. Vector and other tensor fields are denoted by bold face symbols. Spacetime coordinate components are denoted by Greek letters from the middle alphabet, $\mu, \nu, \ldots$ while the indices $\alpha, \beta, \ldots$ label orthonormal tetrad fields. Given a twice covariant tensor field ${\bf T}$ and two vector fields ${\bf X}$ and ${\bf Y}$ we use the notation ${\bf T}({\bf X},{\bf Y}) := T_{\mu\nu} X^\mu Y^\nu$.

\section{Preliminaries}
\label{Sec:Preliminaries}

In this section, we present a brief review of the TB model and state our assumptions on the initial density and radial velocity profiles describing the collapse. Even though these assumptions are identical to the ones made in our previous work~\cite{nOoS11,nOoS14}, we repeat them here for completeness and clarity since they are key to the main results obtained in this work. We also recall some basic properties and state some new results about the behaviour of the metric and fluid fields near the central singularity, which will be needed later in the paper.

With respect to co-moving, synchronous coordinates $(\tau,R,\vartheta,\varphi)$ the spacetime metric ${\bf g}$, four-velocity ${\bf u}$ and energy density $\rho$ of the TB solution are given by
\begin{eqnarray}
{\bf g} &=& -d\tau^2 + \frac{dR^2}{\gamma(\tau,R)^2}
 + r(\tau,R)^2(d\vartheta^2 + \sin^2\vartheta\, d\varphi^2),\qquad
\gamma(\tau,R) := \frac{\sqrt{1 + 2E(R)}}{r'(\tau,R)},
\label{Eq:MetricSol}\\
{\bf u} &=& \frac{\partial}{\partial\tau}\; , \qquad
\rho(\tau,R) = \rho_0(R)\left( \frac{R}{r(\tau,R)} \right)^2\frac{1}{r'(\tau,R)},
\label{Eq:FluidSol}
\end{eqnarray}
where the areal radius $r(\tau,R)$ is a function of proper time $\tau$ measured by observers which are co-moving with the dust particles and the coordinate $R$ labeling the collapsing dust shells. We fix the labeling such that $R$ coincides with the areal radius of the shell at initial time $\tau=0$, that is, $r(0,R) = R$. Explicit expressions for the function $r(\tau,R)$ can be found in Refs.~\cite{rN86,nOoS11} and will not be required in this work. The prime in $r'(\tau,R)$ denotes the partial derivative of $r(\tau,R)$ with respect to $R$. As long as the dust shells do not cross each other, one has $r'(\tau,R) > 0$.

The solution~(\ref{Eq:MetricSol},\ref{Eq:FluidSol}) is parametrized in terms of the initial density $\rho_0(R)$ and radial velocity $v_0(R)$ which determine the function
\begin{equation}
E(R) = \frac{1}{2} v_0(R)^2 - \frac{m(R)}{R},\qquad
m(R) := 4\pi \int_0^R \rho_0(\bar{R})\bar{R}^2 d\bar{R},\qquad R \geq 0.
\label{Eq:Em}
\end{equation}
Here, $m(R)$ describes the total mass contained inside the dust shell $R$ which is independent of $\tau$ as a consequence of mass conservation. Likewise, $E(R)$ describes the total energy of the dust shell $R$ which is conserved in time. Below, we focus on the case $E < 0$ of bounded collapse, though the particular case of marginally bound collapse $E = 0$ can be obtained by taking appropriate limits, as indicated below.

Instead of $\rho_0(R)$ and $v_0(R)$, is it convenient to parametrize the solution in terms of the quantities
$$
c(R) := \frac{2m(R)}{R^3},\qquad
q(R) := \sqrt{\frac{E(R)}{-\frac{m(R)}{R}}} = \sqrt{1 - \frac{R v_0(R)^2}{2m(R)}},
$$
describing (up to a constant factor) the initial mean density within the dust shell $R$ and the square root of the ratio between the total and initial potential energy. In terms of these quantities, our assumptions in~\cite{nOoS11,nOoS14} can be formulated as follows:
\begin{enumerate}
\item[(i)] $\rho_0$ and $v_0$ have even and odd $C^\infty$-extensions, respectively, on the real axis (regular, smooth initial data),
\item[(ii)] $\rho_0(R) > 0$ for all $0\leq R < R_*$ and $\rho_0(R) = 0$ for $R\geq R_*$ (finite, positive density cloud of radius $R_*$),
\item[(iii)] $c'(R)\leq 0$ for all $R > 0$ (monotonically decreasing mean density),
\item[(iv)] $2m(R)/R < 1$ for all $R > 0$ (absence of trapped surfaces on the initial slice),
\item[(v)] $v_0(R)/R < 0$ for all $R\geq 0$ (collapsing cloud),
\item[(vi)] $(v_0(R)/R)^2 < 2m(R)/R^3$  for all $R\geq 0$ (bounded collapse),
\item[(vii)] $q'(R)\geq 0$ for all $R > 0$ (exclusion of shell-crossing singularities),
\item[(viii)] For all $R\geq 0$, we have $q'(R)/R > 0$ whenever $c'(R)/R = 0$ (non-degeneracy condition).
\end{enumerate}
Condition (i) ensures that the functions $c$ and $q$ have even $C^\infty$-extensions on the real axis, and conditions (v) and (vi) imply that $0 < q(R) < 1$ for all $R\geq 0$. Next, as shown in Ref.~\cite{nOoS11}, condition (viii) implies the existence of a null portion of the singularity which is visible at least to local observers\footnote{For studies regarding the structure of the singularity when the non-degeneracy condition (viii) is violated, see for example~\cite{dC84,rN86,iD98,bNfM01,rG06,cU96}.}. Explicit four-parameter families of initial data $(\rho_0(R),v_0(R))$ satisfying all of these conditions with the exception of smoothness of $\rho_0$ and $v_0$ at the surface of the cloud (where these functions are only continuous) have been constructed in Ref.~\cite{nOoS11} [see Eq. (28)] and in Ref.~\cite{nO12} [see Eq.~(4)]. For calculated conformal diagrams displaying the causal structure of the spacetime described by the metric in Eq.~(\ref{Eq:MetricSol}) under the above assumptions (i)--(viii) we refer the reader to Ref.~\cite{nOoS11}. The common feature is the presence of a shell-focusing curvature singularity which forms at the centre $R = 0$ of the cloud after some finite proper time $\tau_s^0$. The first light ray emanating from this singularity describes the Cauchy horizon. Depending on whether the Cauchy horizon intersects the surface of the cloud before or after the apparent horizon, the central singularity is globally naked or hidden inside a black hole. Sufficient conditions for the central singularity to be globally naked are found in Refs.~\cite{dC84,nOoS11}, and this is the case we will be focusing on in this work.

In the remainder of this section we provide a list of explicit expressions for the functions $\gamma$, $\rho$, $r'$ and other relevant functions and their behaviour in the vicinity of the central singularity that will be needed in the following sections. Like in our previous work, we express these quantities in terms of the local radial coordinates $(y,R)$ instead of $(\tau,R)$, where $y$ is defined by
$$
y := \sqrt{\frac{r}{R}}\in [0,1].
$$
In these coordinates, the spacetime manifold $M$ can be characterized by those points $p$ corresponding to $R\geq 0$ and $0 < y \leq 1$ and arbitrary angles $(\vartheta,\varphi)$. The first singular point is characterized by $(y,R) = (0,0)$, and the spacetime region inside the collapsing dust cloud corresponds to the rectangular region $(y,R)\in (0,1)\times (0,R_*)$, with the initial surface and the singularity corresponding to the lines $y=1$ and $y=0$, respectively. Next, we introduce the strictly decreasing function
\begin{equation}
f : [0,1] \to [0,\pi/2], \quad x\mapsto x\sqrt{1-x^2} + \arccos(x),
\label{Eq:fDef}
\end{equation}
which is $C^\infty$-differentiable on the interval $[0,1)$, satisfies $f(0) = \pi/2$, $f(1) = 0$ and has first derivative $f'(x) = -2x^2/\sqrt{1-x^2}$, $0\leq x < 1$, and the functions $g,h: (0,1)\times [0,1)\to\Real$ and $\Lambda: [0,1)\times [0,R_*] \to \Real$ defined as
\begin{eqnarray*}
g(q,y) &:=& \frac{ f(qy) - f(q) }{q^3},\\
h(q,y) &:=& \frac{1}{\sqrt{1-q^2}} - \frac{y^3}{\sqrt{1- q^2y^2}} - \frac{3}{2}g(q,y),\\
\Lambda(y,R) &:=& 2\frac{q'(R)}{Rq(R)} h(q(R),y) - \frac{c'(R)}{2Rc(R)} g(q(R),y).
\end{eqnarray*}
According to Lemma 1 of Ref.~\cite{nOoS11} these functions are strictly positive and $C^\infty$-differentiable on their domain. Note that in the time-symmetric case $q=1$ we have $g(q,y) = f(y)$ and $\Lambda(y,R) = -f(y)c'(R)/[2Rc(R)]$ and the function $h$ is void. In the marginally bound case $q=0$, $g(q,y)$ should be replaced with $g(q,y) = 2(1-y^3)/3$ and in this case $\Lambda(y,R) = -(1-y^3)c'(R)/[3Rc(R)]$. In terms of the coordinates $(y,R)$ the metric coefficient $\gamma$ in Eq.~(\ref{Eq:MetricSol}) and the energy density $\rho$ defined in Eq.~(\ref{Eq:FluidSol}) are:
\begin{eqnarray}
\gamma(y,R) &=& \frac{\sqrt{1 - R^2 q(R)^2 c(R)}}{r'(y,R)},
\label{Eq:gamma}\\
\rho(y,R) &=& \frac{\rho_0(R)}{y^4 r'(y,R)},
\label{Eq:rho}
\end{eqnarray}
where $\rho_0(R) = 2m'(R)/(8\pi R^2) = [R^3 c(R)]'/(8\pi R^2)$ and
\begin{equation}
r'(y,R) = y^2\left( 1 + \frac{R^2}{y^3}\sqrt{1 - q(R)^2y^2} \Lambda(y,R) \right).
\label{Eq:rprime}
\end{equation}
The proper time coordinate $\tau$ as a function of $(y,R)$ is given by
\begin{equation}
\tau(y,R) = \frac{g(q(R),y)}{\sqrt{c(R)}},
\label{Eq:tau}
\end{equation}
and consequently, the collapse time for the dust shell labelled by $R$ is
\begin{equation}
\tau_s(R) := \tau(0,R) = \frac{\frac{\pi}{2} - f(q(R))}{\sqrt{c(R)}q(R)^3},\qquad R\geq 0.
\label{Eq:taus}
\end{equation}
Notice that
\begin{equation}
\frac{\partial\tau(y,R)}{\partial R} = \frac{R}{\sqrt{c(R)}}\Lambda(y,R) > 0, \qquad R > 0,
\label{Eq:dtau}
\end{equation}
which implies, in particular, that $\tau_s(\cdot)$ is a monotonically increasing function. In terms of the original radial coordinates $(\tau,R)$ the manifold $M$ can be characterized by those points with $R\geq 0$ and $0\leq \tau < \tau_s(R)$ and arbitrary angles $(\vartheta,\varphi)$. The first singular point occurs at the centre $R = 0$ at proper time $\tau_s^0 := \tau_s(0)$. We will need the following relation between $\tau_s(R) - \tau$ and $y$:

\begin{lemma}
\label{Lem:Bounds}
Consider a point $p\in M$ with corresponding local radial coordinates $(\tau,R)$ and $(y,R)$. Then the following inequality holds:
\begin{equation}
\frac{2}{3} \frac{1}{\tau_s(R) - \tau} \leq 
\frac{\sqrt{c(R)}}{y^3} \leq \frac{\pi}{2} \frac{1}{\tau_s(R) - \tau}.
\end{equation}
\end{lemma}

\proof According to Eq.~(\ref{Eq:tau}) and the definition of the function $g$ we have
$$
\tau = \frac{f(q(R)y) - f(q(R))}{\sqrt{c(R)} q(R)^3},
$$
with the function $f$ defined in Eq.~(\ref{Eq:fDef}). It is not difficult to verify that this function satisfies the inequalities
$$
\frac{\pi}{2} - \frac{2}{3}x^3 \geq f(x) \geq 
\frac{\pi}{2} (1 - x^3) \quad \mbox{for} \quad 0 \leq x \leq 1,
$$
which yields
$$
\frac{\frac{\pi}{2} - \frac{2}{3}(q(R)y)^3 - f(q(R))}{\sqrt{c(R)} q(R)^3} 
\geq \tau \geq  \frac{\frac{\pi}{2} - \frac{\pi}{2}(q(R)y)^3 - f(q(R)) }{\sqrt{c(R)} q(R)^3}.
$$
Taking into account Eq.~(\ref{Eq:taus}) the statement follows immediately.
\qed

For later use, we will also require the equations
\begin{eqnarray}
\frac{\dot{r}}{r} &=& -\frac{\sqrt{c(R)}}{y^3}\sqrt{1 - q(R)^2 y^2},
\label{Eq:rdot}\\
\frac{\dot{\gamma}}{\gamma} &=& \frac{\sqrt{c}}{y}\frac{H(y,R)}{r'(y,R)},
\label{Eq:gammadot}
\end{eqnarray}
where the dot denotes partial differentiation with respect to $\tau$, and where the smooth function $H: (0,1)\times (0,R_*)\to \Real$ defined in Ref.~\cite{nOoS14} is given by
\begin{equation}
H(y,R) := \frac{1}{\sqrt{1-q(R)^2y^2}}
\left[ 1 + R\frac{c'(R)}{2c(R)} - q(R)^2 y^2\left( 1 + R\frac{c'(R)}{2c(R)} + R\frac{q'(R)}{q(R)} \right) \right]
 -\frac{1}{2}\frac{R^2}{y^3}\Lambda(y,R).
\end{equation}

\begin{figure}[h!]
\begin{center}
\includegraphics[width=9cm]{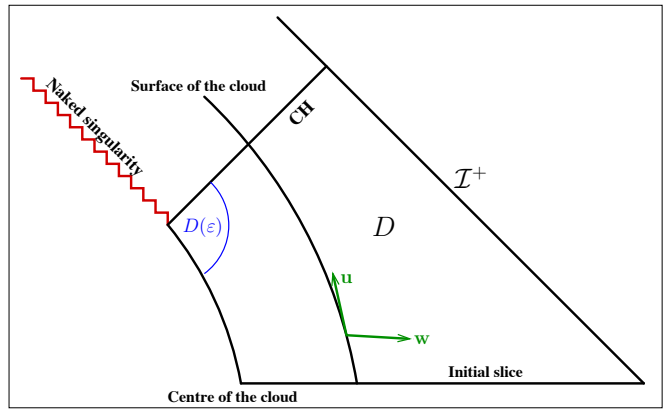}
\end{center}
\caption{\label{Fig:D_epsilon} Conformal diagram illustrating the maximal development $D$ of the initial slice and the small subsets $D(\varepsilon)$ close to the central singularity. Here ``CH'' denotes the Cauchy horizon, and ${\bf u}$ and ${\bf w}$ form an orthonormal basis of radial vector fields. Reproduced from~\cite{nOoS14}. IOP Publishing Ltd. All rights reserved.}
\end{figure}

As mentioned before, in the following, we only consider the case of a globally naked singularity. More specifically, we focus our attention on the region $D$ of spacetime describing the maximal development of the initial surface $\tau=0$, see figure~\ref{Fig:D_epsilon}. In terms of the coordinates $(y,R)$ this region is described by
$$
D := \{ p\in M : R\geq 0, y_{CH}(R) < y \leq 1 \},
$$
where the function $y_{CH}(R)$ describes the location of the Cauchy horizon. For $0 < R < \delta$ small enough it was shown in Proposition~2 of Ref.~\cite{nOoS11} that $y_{CH}(R)$ has the form
\begin{equation}
y_{CH}(R)^3 = \frac{3\Lambda_0}{4} R^2\left[ 1 + \zeta(R^{1/3}) \right],\qquad
R\in [0,\delta),
\label{Eq:yCH}
\end{equation}
where $\Lambda_0 := \Lambda(0,0) > 0$ and $\zeta: [0,\delta^{1/3})\to\Real$ is a $C^\infty$-function satisfying $\zeta(0) = 0$ and $\zeta'(0) = -3(3\Lambda_0/4)^{-1/3}\sqrt{c(0)}/2$. Besides the spacetime region $D$, we also consider as in~\cite{nOoS14} for each $\varepsilon \in (0,1)$ the small spacetime regions $D(\varepsilon)\subset D$ (see figure~\ref{Fig:D_epsilon}) near the central singularity defined as
\begin{equation}
D(\varepsilon) := \{p\in M : 0 \leq R \leq R(\varepsilon), y_{CH}(R) < y \leq y(\varepsilon) \},
\label{Eq:DepsDef}
\end{equation}
where the functions $R(\varepsilon) \in (0,\delta)$ and $y(\varepsilon) := y_{CH}(R(\varepsilon))$ are chosen such that they converge to zero when $\varepsilon\to 0$ and such that
\begin{enumerate}
\item[(i)] $\zeta(R^{1/3}) \geq -\varepsilon$ for all $0\leq R \leq R(\varepsilon)$,
\item[(ii)] $\Lambda(y,R)\leq (1 + \varepsilon)\Lambda_0$ for all $0\leq R\leq R(\varepsilon)$ and $0\leq y\leq y(\varepsilon)$.
\end{enumerate}

It follows for each $p\in D(\varepsilon)$ that
\begin{equation}
0 \leq \frac{R^2}{y^3}\leq \frac{R^2}{y_{CH}(R)^3}
 \leq \frac{4}{3\Lambda_0}\frac{1}{1 + \zeta(R^{1/3})}
 \leq \frac{4}{3\Lambda_0}\frac{1}{1 - \varepsilon}.
\label{Eq:R2y3Estimate}
\end{equation}

With these observations we can easily prove the following result from Eqs.~(\ref{Eq:gamma},\ref{Eq:rho},\ref{Eq:rprime},\ref{Eq:rdot},\ref{Eq:gammadot}):

\begin{lemma}[see also Lemmata~1 and~3 in~\cite{nOoS14}]
\label{Lem:Estimates}
Let $\varepsilon\in (0,1)$. There are constants $C >  c > 0$ such that for all $p\in D(\varepsilon)$ the following inequalities hold:
\begin{equation}
\frac{c}{y^2} \leq \gamma \leq \frac{C}{y^2},\qquad
\frac{c}{y^3} \leq -\frac{\dot{r}}{r} \leq \frac{C}{y^3},\qquad
\frac{c}{y^3} \leq \frac{\dot{\gamma}}{\gamma} \leq \frac{C}{y^3},\qquad
\frac{c}{y^6} \leq \rho \leq \frac{C}{y^6}.
\end{equation}
\end{lemma}
Therefore, $\gamma$, $\dot{r}/r$, $\dot{\gamma}/\gamma$ and $\rho$ diverge like $1/y^2$, $1/y^3$, $1/y^3$ and $1/y^6$, respectively, when the central singularity is approached from within region $D(\varepsilon)$.

Finally, we introduce the vector fields ${\bf u}$ and ${\bf w}$, where ${\bf u} := \partial_\tau$ is the four-velocity of free-falling observers co-moving with the collapsing dust particles, and ${\bf w} := \gamma(\tau,R)\partial_R$ is the unit outward radial vector orthogonal to ${\bf u}$, see figure~\ref{Fig:D_epsilon}. These fields may be completed to a (local) orthonormal frame $\{ {\bf u}, {\bf w}, {\bf e}_{\vartheta}, {\bf e}_{\varphi} \}$, which in turn naturally determines a Newman-Penrose null tetrad $\{ {\bf k}, {\bf l}, {\bf m}, {\bf m}^* \}$ with in- and outgoing null vector fields ${\bf l} := {\bf u} - {\bf w}$ and ${\bf k} := {\bf u} + {\bf w}$, and ${\bf m} = {\bf e}_{\vartheta} + i{\bf e}_{\varphi}$ a complex null vector with complex conjugate ${\bf m}^*$, such that $g_{\mu\nu} = -k_{(\mu} l_{\nu)} + m_{(\mu} m_{\nu)}^*$. With respect to this tetrad, the Weyl tensor ${\bf C}$ associated with the metric~(\ref{Eq:MetricSol}) can be decomposed into the complex Newman-Penrose scalars $\Psi_{-2}$, $\Psi_{-1}$, $\Psi_0$, $\Psi_1$ and $\Psi_2$. Because of the spherical symmetry of the metric~(\ref{Eq:MetricSol}), the only nonvanishing scalar is
\begin{equation}
\Psi_0 := C_{\alpha\beta\gamma\delta } k^\alpha m^\beta (m^*)^\gamma l^\delta
 = \frac{16\pi}{3}\left( \rho - \frac{m}{\frac{4\pi}{3} r^3} \right).
\end{equation}
Up to a constant factor, $\Psi_0$ is equal to the difference between the density and the mean density of the dust cloud. Using Eqs.~(\ref{Eq:FluidSol},\ref{Eq:Em}) we can rewrite this as
\begin{equation}
\Psi_0(y,R) = \frac{2c(R)}{3y^6}
\frac{ R\frac{c'(R)}{c(R)} - 3\frac{R^2}{y^3}\sqrt{1 - q(R)^2 y^2}\Lambda(y,R)}
{1 + \frac{R^2}{y^3}\sqrt{1 - q(R)^2 y^2}\Lambda(y,R)}.
\end{equation}
Note that $\Psi_0(y,0) = 0$ for all $0 < y\leq 1$, reflecting the fact that for $R\to 0$ the mean density converges to $\rho$. However, $\Psi_0$ diverges as $1/y^6$ as the central singularity is approached along the Cauchy horizon: indeed, it follows from the above remarks that
\begin{equation}
\lim\limits_{R\to 0} y_{CH}(R)^6\Psi_0(y_{CH}(R),R) = -\frac{8c(0)}{7} < 0.
\label{Eq:Psi0Bound}
\end{equation}

\section{Energy estimates}
\label{Sec:EnergyEstimates}

In this section, we consider a (linear or nonlinear) test field $\Phi$ propagating on the region $D$ of a TB spacetime with a globally naked singularity (see figure~\ref{Fig:D_epsilon}). We assume that $\Phi$ is governed by a well-posed initial value problem on $D$ and that $\Phi$ possesses a stress-energy tensor ${\bf T} = T_{\mu\nu} dx^\mu\otimes dx^\nu$ which, as a consequence of the equations of motion, is divergence-free and satisfies the dominant energy condition.

The main result of this section is to show that a suitable energy norm of $\Phi$ (which is constructed from ${\bf T}$) is uniformly bounded on $D$, provided the initial data is sufficiently smooth and regular at $R = 0$. For the sake of clarity of the presentation, we split this section into four subsections. We first prove some basic results which are direct consequences of the dominant energy condition. Then, we derive a standard balance equation from which the energy estimates are obtained. Next, we estimate the source term in the balance equation. This is the most technical part of the argument, so some of the proofs will be given in an appendix. Finally, we formulate our main result. Specific applications are discussed in the next section. Although the material in the first two subsections is rather standard (see for instance~\cite{HawkingEllis-Book}), we include it for completeness and in order to introduce the required notation for the energy, flux and source quantities.

\subsection{Dominant energy condition and consequences}

The dominant energy condition states that for any future-directed timelike vector field ${\bf X}$, the vector field ${\bf J}_{\bf X}$ whose components are $J_{\bf X}^\mu := -T^\mu{}_\nu X^\nu$ should be future-directed timelike or null. The following result will be used repeatedly in this work:

\begin{lemma}
\label{Lem:DEC}
Let ${\bf T}$ be a stress-energy tensor satisfying the dominant energy condition, let $\{ {\bf e}_0, {\bf e}_1,{\bf e}_2, {\bf e}_3 \}$ be a (local) orthonormal frame, ${\bf e}_0$ being future-directed timelike, and let ${\bf X}$ be a future-directed causal vector field. Then,
\begin{equation}
|{\bf T}({\bf e}_\alpha,{\bf X})| \leq {\bf T}({\bf e}_0,{\bf X}),
\qquad \alpha = 0,1,2,3.
\label{Eq:DEC1}
\end{equation}
Furthermore,
\begin{equation}
|{\bf T}({\bf e}_\alpha,{\bf e}_\beta)| \leq {\bf T}({\bf e}_0,{\bf e}_0),
\qquad \alpha,\beta = 0,1,2,3,
\label{Eq:DEC2}
\end{equation}
that is, all the orthonormal components of ${\bf T}$ are bounded in magnitude by the energy density ${\bf T}({\bf e}_0,{\bf e}_0)$ measured by the observer with four-velocity ${\bf e}_0$.
\end{lemma}

\proof According to the dominant energy condition the vector field ${\bf J}_{\bf X} := -T^\alpha{}_\beta X^\beta e_\alpha$ is future-directed causal. Therefore,
$$
{\bf T}({\bf e}_0,{\bf X}) = -{\bf g}({\bf e}_0,{\bf J}_{\bf X}) \geq 0,
$$
and for $\alpha = 1,2,3$,
$$
|{\bf T}({\bf e}_\alpha,{\bf X})| = |{\bf g}({\bf e}_\alpha,{\bf J}_{\bf X})| 
 \leq |{\bf g}({\bf e}_0,{\bf J}_{\bf X})| = {\bf T}({\bf e}_0,{\bf X}),
$$
which proves the first assertion, Eq.~(\ref{Eq:DEC1}). For $\alpha = 0,1,2,3$ and $\beta = 0$ the second assertion is a direct consequence of the first one, so it only remains to prove Eq.~(\ref{Eq:DEC2}) for $\alpha,\beta = 1,2,3$. To this purpose we introduce ${\bf X} = {\bf e}_0 \pm {\bf e}_\beta$ into Eq.~(\ref{Eq:DEC1}), obtaining $| T_{\alpha 0} \pm T_{\alpha\beta} | \leq T_{00} \pm T_{0\beta}$ for all $\alpha,\beta=1,2,3$. This yields the three inequalities
\begin{eqnarray*}
T_{\alpha\beta} + T_{0\alpha} &\leq& T_{00} + T_{0\beta},\\
-T_{\alpha\beta} + T_{0\alpha} &\leq& T_{00} - T_{0\beta},\\
-T_{\alpha\beta} - T_{0\alpha} &\leq& T_{00} + T_{0\beta},
\end{eqnarray*}
for arbitrary $\alpha,\beta = 1,2,3$. Symmetrizing the first one with respect to $\alpha$ and $\beta$ yields $T_{\alpha\beta}\leq T_{00}$, and summing the last two gives $-T_{\alpha\beta} \leq T_{00}$, which concludes the proof of the lemma.
\qed

\subsection{Balance laws}
\label{Sub:BalanceEqn}

In this subsection, we discuss the balance laws which are obtained from the contraction of the stress-energy tensor ${\bf T}$ and a future-directed timelike smooth vector field ${\bf X}$ on $D$ (which is left unspecified for the moment). Since ${\bf T}$ is divergence-free, the associated current density $J_{\bf X}^\mu = -T^\mu{}_\nu X^\nu$ satisfies
\begin{equation}
\nabla_\mu J_{\bf X}^{\mu} = S_{\bf X},\qquad
S_{\bf X} := -T^{\mu \nu} \nabla_{\mu} X_{\nu}.
\label{Eq:DivergenceLaw}
\end{equation}
The source term $S_{\bf X}$ vanishes if ${\bf X}$ is a Killing vector field. However, in our case we cannot assume that ${\bf X}$ is Killing since the spacetime in the interior of the cloud is dynamical.

\begin{figure}[h!]
\begin{center}
\includegraphics[height=7.cm,width=8.4cm]{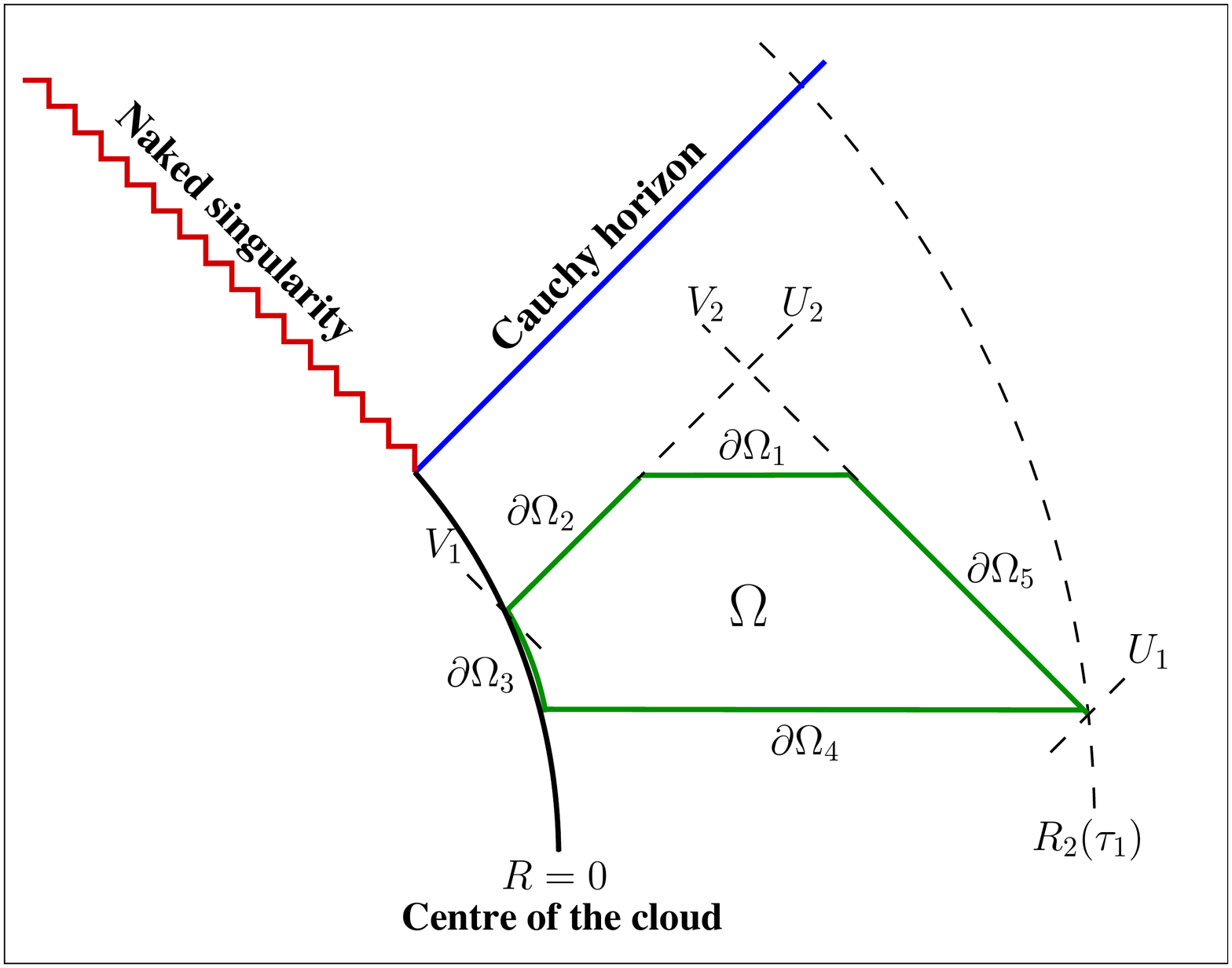}
\hspace{-0.1cm}
\includegraphics[height=7.cm,width=9.4cm]{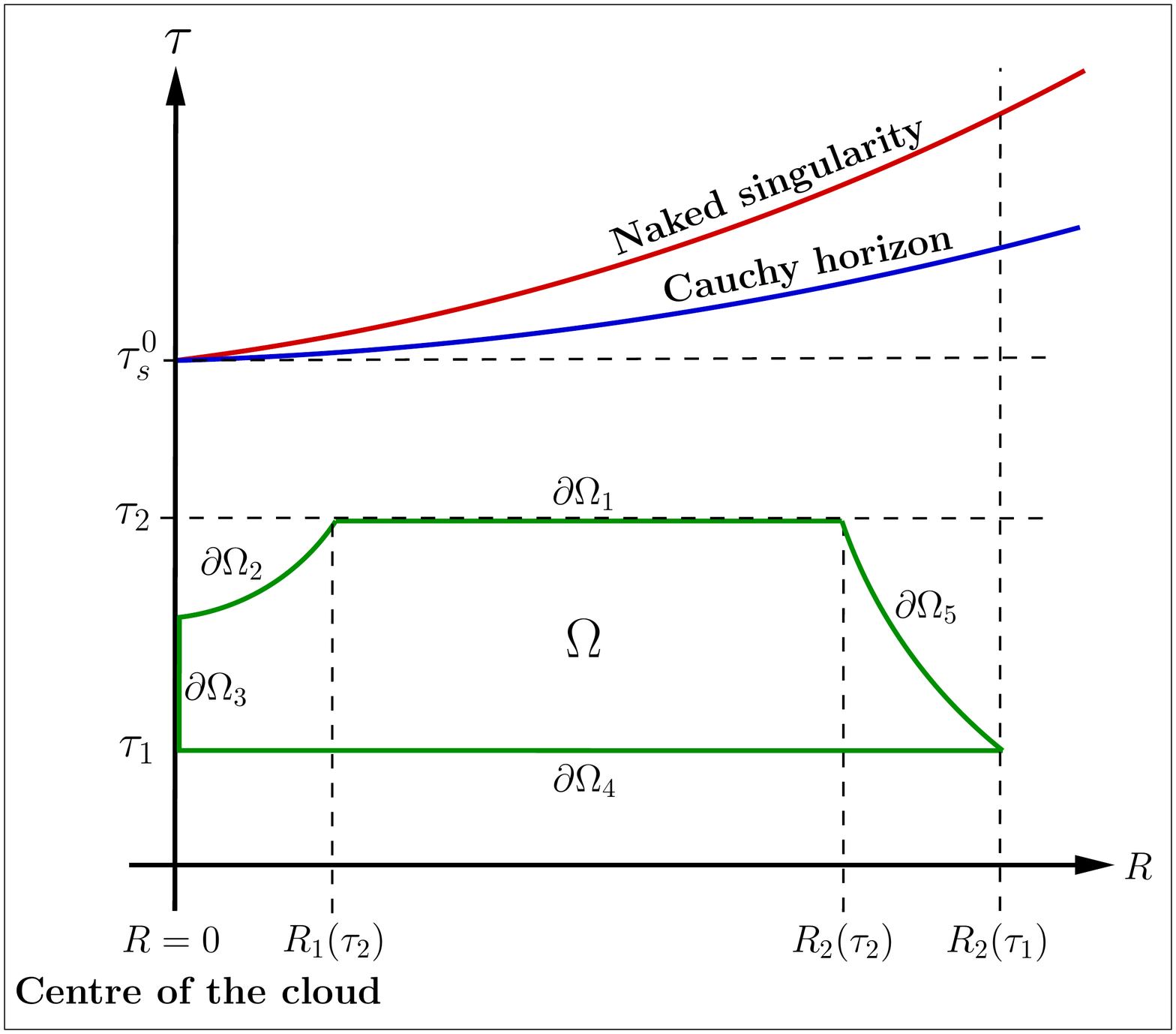}
\end{center}
\caption{\label{Fig:Estimate_region1} Illustration of the particular domain $\Omega\subset D$ used for the balance equation~(\ref{Eq:Gauss_applied}). Left panel: Conformal diagram based on the coordinate system $(T,X) = (V + U,V-U)/2$ (see Ref.~\cite{nOoS11} for more details). Right panel: Diagram based on the co-moving synchronous coordinates $(\tau,R)$. In both diagrams, the angular coordinates are suppressed.}
\end{figure}

A balance law is obtained by integrating both sides of Eq.~(\ref{Eq:DivergenceLaw}) over a compact domain $\Omega\subset D$. For our purposes, we choose $\Omega$ as illustrated in figure~\ref{Fig:Estimate_region1}. $\Omega$ is delimited by the spacelike surfaces $\partial\Omega_1$ and $\partial\Omega_4$ of constant $\tau$ and by the rotational-invariant null surfaces $\partial\Omega_2$ and $\partial\Omega_5$ of constant retarded and advances times $U$ and $V$, respectively, see figure~\ref{Fig:Estimate_region1}. Here, $U$ and $V$ are defined by
$$
\dot{U} = -\gamma U',\qquad
\dot{V} = +\gamma V',
$$
and the initial conditions $-U(\tau=0,R) = V(\tau=0,R) = R$, $R\geq 0$, and boundary conditions $U(\tau,0) = V(\tau,0)$, $0\leq \tau < \tau_s^0$. Note that
$$
\left( \frac{\partial}{\partial\tau} - \gamma\frac{\partial}{\partial R}\right)\dot{V}
 = \frac{\dot{\gamma}}{\gamma}\dot{V},
$$
and since $\dot{\gamma}/\gamma$ is regular away from the singularity, it follows that $\dot{V}$ cannot change sign along radial incoming null geodesics. Likewise, $\dot{U}$ cannot change sign along radial outgoing null geodesics, and since $\dot{U}$ and $\dot{V}$ are positive at $\tau=0$ they are positive everywhere in $D$. For the following, we denote by $\Sigma_U$ and $\Sigma_V$ the null hypersurfaces of constant $U$ and $V$, respectively, and by $\Sigma_\tau$ the hypersurfaces of constant $\tau$.

Assuming without loss of generality that $\Omega$ is contained inside a single coordinate chart $(x^\mu)$ (otherwise we use a partition of unity), we obtain upon integration of Eq.~(\ref{Eq:DivergenceLaw}) and using Gauss' theorem in $\Real^4$,
\begin{equation}
\int_{\partial\Omega} J_{\bf X}^{\mu} n_{\mu} \sqrt{|g|} dS
 = \int_\Omega S_{\bf X} \sqrt{|g|} d^4 x,
\label{Eq:Gauss_applied}
\end{equation}
where $(n_\mu)$ is the outward unit normal with respect to the Euclidean metric in $\Real^4$ to $\partial\Omega$, $|g|$ is the absolute value of the determinant of $(g_{\mu\nu})$, and $dS$ is the induced volume element on $\partial\Omega$. We compute the left-hand side of Eq.~(\ref{Eq:Gauss_applied}) for each component of $\partial\Omega$: for $\partial\Omega_1$ and $\partial\Omega_4$ we use the local coordinates $(\tau,R,\vartheta,\varphi)$ for which the metric has the form~(\ref{Eq:MetricSol}) and $\sqrt{|g|} = r^2\gamma^{-1}\sin\vartheta$. The outward unit normals on $\partial\Omega_1$ and $\partial\Omega_4$ have components
$$
\left( n_{\mu}^{(1)} \right) = (1,0,0,0) = -(u_\mu),\qquad
\left( n_{\mu}^{(4)} \right) = (-1,0,0,0) = (u_\mu),
$$
respectively, and hence we obtain
$$
\int_{\partial\Omega_1} J_{\bf X}^{\mu} n_{\mu} \sqrt{|g|} dS
 = \int_{R_1(\tau_2)}^{R_2(\tau_2)} 
 \left. \langle {\bf T}({\bf X},{\bf u}) \rangle_{S^2} r^2 \frac{dR}{\gamma} \right|_{\tau=\tau_2},
\qquad
\int_{\partial\Omega_4} J_{\bf X}^{\mu} n_{\mu} \sqrt{|g|} dS
 = -\int_0^{R_2(\tau_1)} 
 \left. \langle {\bf T}({\bf X},{\bf u}) \rangle_{S^2} r^2 \frac{dR}{\gamma} \right|_{\tau=\tau_1},
$$
where
$$
\langle f \rangle_{S^2} := \int_{S^2} f \sin\vartheta\, d\vartheta d\varphi
$$
denotes (up to a factor $1/4\pi$) the mean value of $f$ over the $2$-sphere $S^2$. As a consequence of Lemma~\ref{Lem:DEC} the quantity $\langle {\bf T}({\bf X},{\bf u}) \rangle_{S^2}$ is nonnegative, and hence
\begin{equation}
{\cal E}_{\bf X}(\tau) := \int_{\Omega\cap \Sigma_\tau} {\bf T}({\bf X},{\bf u})
 = \int_{R_1(\tau)}^{R_2(\tau)} 
 \left. \langle r^2 {\bf T}({\bf X},{\bf u}) \rangle_{S^2} \frac{dR}{\gamma} \right|_{\tau},
\label{Eq:Energy_def}
\end{equation}
defines an \emph{energy-type norm} for the test field on the intersection of the spacelike surface $\Sigma_\tau$ with the domain $\Omega$.

In order to compute the integral over the boundary components $\partial\Omega_2$ and $\partial\Omega_5$ we use the double-null coordinates $(U,V,\vartheta,\varphi)$ with respect to which
$$
{\bf g} = -\frac{1}{\dot{U}\dot{V}} dU dV + r^2(d\vartheta^2 + \sin^2\vartheta\, d\varphi^2),
$$
and $\sqrt{|g|} = r^2\sin\vartheta/(2\dot{U}\dot{V})$. The outward unit normals on $\partial\Omega_2$ and $\partial\Omega_5$ have the components
$$
\left( n_{\mu}^{(2)} \right) = (1,0,0,0) = -\dot{U}(k_\mu),\qquad
\left( n_{\mu}^{(5)} \right) = (0,1,0,0) = -\dot{V}(l_\mu),
$$
respectively, and hence we obtain
$$
\int_{\partial\Omega_2} J_{\bf X}^{\mu} n_{\mu} \sqrt{|g|} dS = {\cal F}_{\bf X}^-(U_2),\qquad
\int_{\partial\Omega_5} J_{\bf X}^{\mu} n_{\mu} \sqrt{|g|} dS = {\cal F}_{\bf X}^+(V_2),
$$
where the \emph{flux integrals} ${\cal F}_{\bf X}^\pm$ are defined by
\begin{eqnarray}
{\cal F}_{\bf X}^-(U) &:=& \int_{\Sigma_U \cap \Omega}{\bf T} ({\bf X}, {\bf k})
 = \int_{V_1}^{V_2} 
 \left.\langle r^2 {\bf T} ({\bf X}, {\bf k}) \rangle_{S^2} \frac{dV}{\dot{V}} \right|_U,
\label{Eq:Fluxes_def1}\\
{\cal F}_{\bf X}^+(V) &:=& \int_{\Sigma_V \cap \Omega}{\bf T} ({\bf X}, {\bf l})
 = \int_{U_1}^{U_2} 
 \left. \langle r^2 {\bf T} ({\bf X}, {\bf l}) \rangle_{S^2} \frac{dU}{\dot{U}} \right|_V.
\label{Eq:Fluxes_def2}
\end{eqnarray}
These fluxes are nonnegative as a consequence of Lemma~\ref{Lem:DEC}. With these observations, Eq.~(\ref{Eq:Gauss_applied}) can be reformulated in the following way:

\begin{proposition}[Balance law]
\label{Prop:Balance}
Let ${\bf T}$ be a smooth, covariant symmetric tensor field on $(M,{\bf g})$ which satisfies the divergence law $\nabla^\mu T_{\mu\nu} = 0$, and let ${\bf X}$ be a smooth vector field on $(M,{\bf g})$. Let $\Omega\subset D$ be a compact region of $D$ as described in figure~\ref{Fig:Estimate_region1}, and let ${\cal E}_{\bf X}(\tau)$, ${\cal F}_{\bf X}^-(U)$ and ${\cal F}_{\bf X}^+(V)$ be defined by Eq.~(\ref{Eq:Energy_def},\ref{Eq:Fluxes_def1},\ref{Eq:Fluxes_def2}). Then,
\begin{equation}
{\cal E}_{\bf X}(\tau_2) + {\cal F}_{\bf X}^-(U_2) + {\cal F}_{\bf X}^+(V_2) 
 = {\cal E}_{\bf X}(\tau_1) + \int_\Omega S_{\bf X} \sqrt{|g|} d^4x,
\label{Eq:energy_balance_tau_R}
\end{equation}
where $S_{\bf X} = -T^{\mu\nu}\nabla_\mu X_\nu$. Furthermore, if ${\bf T}$ satisfies the dominant energy condition and ${\bf X}$ is future-directed timelike, then ${\cal E}_{\bf X}(\tau)\geq 0$, ${\cal F}_{\bf X}^-(U)\geq 0$, and ${\cal F}_{\bf X}^+(V)\geq 0$.
\end{proposition}

The main result in the next subsection is to show that for a suitable choice of ${\bf X}$ the volume integral over $S_{\bf X}$ can be bounded from above as
\begin{equation}
\int_\Omega S_{\bf X} \sqrt{|g|} d^4x \leq 
\int_{\tau_1}^{\tau_2} \alpha(\tau) {\cal E}_{\bf X}(\tau) d\tau,
\label{Eq:SXEstimate}
\end{equation}
with $\alpha: (0,\infty)\to \Real$ a nonnegative, integrable function. As a consequence of Proposition~\ref{Prop:Balance} we then obtain the following energy estimate:
\begin{equation}
{\cal E}_{\bf X}(\tau_2) \leq  {\cal E}_{\bf X}(\tau_1) 
 + \int_{\tau_1}^{\tau_2}  \alpha(\tau) {\cal E}_{\bf X}(\tau) d\tau,
\quad \tau_2 > \tau_1 > 0,
\label{Eq:EXEstimate}
\end{equation}
from which a uniform energy bound is obtained using Gronwall's inequality~\cite{Evans-Book}.

\subsection{Estimates for the source term $S_{\bf X}$}
\label{Sub:Source}

The goal of this subsection is to establish the bound~(\ref{Eq:SXEstimate}) for some suitable future-directed timelike vector field ${\bf X}$. The key question is the choice for ${\bf X}$ which is a priori not obvious since our spacetime is not stationary inside the cloud.

Given the four-velocity ${\bf u}$ of the dust particles, a natural choice consists in ${\bf X} = {\bf u}$. A short computation reveals that
\begin{equation}
\nabla_a u_b = -\frac{\dot{\gamma}}{\gamma} w_a w_b, \quad
\nabla_a u_B = \nabla_B u_a = 0, \quad 
\nabla_A u_B = \frac{\dot{r}}{r} g_{AB},
\label{Eq:gradu}
\end{equation}
where here $(x^a) = (\tau,R)$ and $(x^A) = (\vartheta,\varphi)$ refer to radial and angular coordinates, respectively, and ${\bf w}$ is the unit outward radial vector orthogonal to {\bf u} defined in section~\ref{Sec:Preliminaries}. Accordingly, we find
$$
S_{\bf u} = \frac{\dot{\gamma}}{\gamma}{\bf T}({\bf w},{\bf w}) 
 - \frac{\dot{r}}{r} g^{AB} T_{AB},
$$
and Lemma~\ref{Lem:DEC} leads to the estimate
$$
\int_\Omega S_{\bf u} \sqrt{|g|} d^4x 
 = \int_{\tau_1}^{\tau_2}
  \left( \int_{\Sigma_\tau\cap \Omega} S_{\bf u} \right) d\tau
 \leq \int_{\tau_1}^{\tau_2} \int_{\Sigma_\tau\cap \Omega}
\left( \left| \frac{\dot{\gamma}}{\gamma} \right| + 2\left| \frac{\dot{r}}{r} \right| \right)
{\bf T}({\bf u},{\bf u}) d\tau
\leq \int_{\tau_1}^{\tau_2} \alpha(\tau) {\cal E}_{\bf u}(\tau) d\tau
$$
with the function
$$
\alpha(\tau) := \sup\limits_{R > R_1(\tau)} \left[
\left( \left| \frac{\dot{\gamma}}{\gamma} \right| + 2\left| \frac{\dot{r}}{r} \right| \right)(\tau,R) \right], \qquad \tau \geq 0.
$$
However, the estimates in Lemma~\ref{Lem:Estimates} imply that both terms $\dot{\gamma}/\gamma$ and $\dot{r}/r$ diverge like $1/y^3$ in the region $D(\varepsilon)$ close to the central singularity. On the other hand, according to Lemma~\ref{Lem:Bounds}, the term $1/y^3$ diverges like $(\tau_s(R) - \tau)^{-1}$ as the central singularity at $\tau = \tau_s^0$ is approached. Consequently, it follows that the integral of $\alpha$ diverges logarithmically as $\tau\to \tau_s^0$, and it does not look like we can obtain a uniform energy bound with the choice ${\bf X} = {\bf u}$. The reason for the fast divergence of the source term $S_{\bf u}$ relies in the fact that the geodesic congruence describing the free-falling dust particles with four-velocity ${\bf u}$ converges at the central singularity, leading to a fast-diverging (negative) expansion and shear which contribute to $S_{\bf u}$.

For this reason, we choose the vector field ${\bf X}$ differently. Instead of choosing it to be tangent to the surfaces of the collapsing dust shells of constant $R$, we choose ${\bf X}$ to be tangent to the surfaces of constant \emph{areal} radius $r$. As we will see shortly, such a choice leads to a slower divergence of the gradients $\nabla_\mu X_\nu$ at the central singularity. An explicit and natural choice is given by the Kodama vector field~\cite{hK80}, which in our notation reads
\begin{equation}
{\bf X} :=  \gamma r' {\bf u} - \dot{r} {\bf w}.
\label{Eq:X_def}
\end{equation}

\begin{lemma}
\label{Lem:XProperties}
The vector field ${\bf X}$ defined in Eq.~(\ref{Eq:X_def}) satisfies the following properties:
\begin{enumerate}
\item[(a)] ${\bf X}[r] = 0$,
\item[(b)] $-{\bf g}({\bf X},{\bf X}) = 1 - \frac{2m(R)}{r} = 1 - \frac{R^2 c(R)}{y^2}$,
\item[(c)] $-{\bf g}({\bf X},{\bf u}) = \sqrt{1 + 2E(R)} = \sqrt{1 - R^2 q(R)^2 c(R)}$,
\item[(d)] $\nabla_a X_b = -\frac{m}{r^2} u_a w_b 
 + \left( \frac{m}{r^2} - 4\pi r\rho \right) w_a u_b,\qquad
\nabla_a X_B = \nabla_A X_b = 0,\qquad \nabla_A X _B = 0$.
\end{enumerate}
\end{lemma}

\proof (a), (b) and (c) follow from a direct calculation based on Eqs.~(\ref{Eq:MetricSol}) and (\ref{Eq:gamma}). As for (d), one uses the identities
\begin{equation}
\nabla_a w_b = -\frac{\dot{\gamma}}{\gamma} w_a u_b, \quad
\nabla_a w_B = \nabla_B w_a = 0, \quad 
\nabla_A w_B = \frac{\gamma r'}{r} g_{AB},
\label{Eq:gradw}
\end{equation}
and finds
$$
\nabla_a X_b = \ddot{r} u_a w_b + \left[ \frac{\dot{\gamma}}{\gamma} \dot{r} + \gamma (\gamma r')' \right] w_a u_b,\qquad
\nabla_a X_B = \nabla_A X_b = 0,\qquad \nabla_A X _B = 0.
$$
Finally, using the equations of motion $\ddot{r} = -m/r^2$ and $\dot{r}^2/2 = E + m/r$, one obtains
$$
\frac{\dot{\gamma}}{\gamma} \dot{r} + \gamma (\gamma r')'
 = -\dot{r}\frac{\dot{r}'}{r'} + \frac{\sqrt{1 + 2E}}{r'}\left( \sqrt{1 + 2E} \right)'
 = -\frac{1}{r'} \left( \frac{1}{2}\dot{r}^2 \right)' + \frac{E'}{r'} = \frac{m}{r^2} - \frac{m'}{r r'},
$$
from which the statement follows.
\qed

As a consequence of property (a), ${\bf X}$ is tangent to the hypersurfaces of constant areal radius $r$, as wished. Next, property (b) implies that ${\bf X}$ is timelike in the region exterior to the apparent horizon $r > 2m$ in which $D$ is contained, and property (c) then shows that ${\bf X}$ is also future-directed in this region. Finally, property (d) implies that the symmetric part of $\nabla_\mu X_\nu$ which appears in the source term $S_{\bf X}$ has the following simple form:
\begin{equation}
S_{\bf X} = 4\pi r\rho {\bf T}({\bf u},{\bf w}).
\label{Eq:Source}
\end{equation}
A short computation reveals that
\begin{equation}
r\rho = \frac{1}{y^{5/2}} \left( \frac{R^2}{y^3} \right)^{1/2}\frac{\rho_0(R)}{1 + {\frac{R^2}{y^3}\sqrt{1-q(R)^2y^2} \Lambda(y,R)}}. 
\label{Eq:SX}
\end{equation}
An important observation is the fact that the coefficient $r\rho$ in the source term diverges only as $1/y^{5/2}$ in the region $D(\varepsilon)$ where the estimate~(\ref{Eq:R2y3Estimate}) holds, whereas the analogous coefficient diverged as $1/y^3$ with the previous choice ${\bf X} = {\bf u}$. As we show next, this small gain in the exponent characterizing the divergence of the source term ($5/2$ instead of $3$) is sufficient to obtain a uniform bound for the energy ${\cal E}_{\bf X}$. We also note that the right-hand side of Eq.~(\ref{Eq:Source}) vanishes in the exterior of the cloud. This is expected since in the Schwarzschild region ${\bf X}$ coincides with the timelike Killing vector field.

The key result for obtaining the uniform energy bound is the following:

\begin{proposition}
\label{Prop:alpha}
Let ${\bf X}$ be the vector field defined in Eq.~(\ref{Eq:X_def}) which is future-directed timelike on $D$. Consider the compact domain $\Omega\subset D$ shown in figure~\ref{Fig:Estimate_region1}. Then, there is a nonnegative integrable function $\alpha: [0,\infty)\to\Real$ such that for all $\tau_2\geq \tau_1\geq 0$
\begin{equation}
\int_\Omega S_{\bf X} \sqrt{|g|} d^4x \leq 
\int_{\tau_1}^{\tau_2} \alpha(\tau) {\cal E}_{\bf X}(\tau) d\tau.
\end{equation}
\end{proposition}

\proof Using Eq.~(\ref{Eq:Source}) we first find
\begin{eqnarray}
\int_\Omega S_{\bf X} \sqrt{|g|} d^4x 
 &=& 4\pi \int_{\tau_1}^{\tau_2} \left( \int_{\Sigma_\tau\cap \Omega}
 r\rho {\bf T}({\bf u},{\bf w}) \right) d\tau
\nonumber \\
&\leq& 4\pi \int_{\tau_1}^{\tau_2} 
\sup\limits_{R_1(\tau) \leq R \leq R_2(\tau)} \left[ (r\rho)(\tau,R) \right] 
\left( \int_{\Sigma_\tau\cap \Omega} |{\bf T}({\bf u},{\bf w})| \right) d\tau.
\label{Eq:First_estimate}
\end{eqnarray}
In order to estimate $|{\bf T}({\bf u},{\bf w})|$ we use the following Lemma which is a consequence of the properties of ${\bf X}$ and Lemma~\ref{Lem:DEC}:

\begin{lemma}
\label{Lem:TXu}
Let $\chi(y,R) := \sqrt{2m(R)/r}$ be the square root of the compactness ratio of the dust shell $R$, and let
\begin{equation}
C_0 := \sup\limits_{\substack{R > 0\\ y_{CH}(R) < y \leq 1}}\frac{1}{1 - \chi(y,R)},
\end{equation}
which is finite due to our assumption that the singularity is globally naked, which means that the region $D$ stays away from the apparent horizon at $r = 2m$.

Then the following estimates are valid on $D$:
\begin{enumerate}
\item[(a)] ${\bf T}({\bf u},{\bf Y}) \leq C_0 {\bf T}({\bf X},{\bf Y})$
for any future-directed causal vector field ${\bf Y}$ on $D$. In particular, ${\bf T}({\bf u},{\bf u}) \leq C_0 {\bf T}({\bf X},{\bf u})$ on $D$.
\item[(b)] $|{\bf T}({\bf u},{\bf w})| \leq C_0 {\bf T}({\bf X},{\bf u})$.
\item[(c)] $|{\bf T}({\bf w},{\bf w})| \leq C_0 {\bf T}({\bf X},{\bf u})$.
\item[(d)] ${\bf T}({\bf u},{\bf X}) \leq 2{\bf T}({\bf u},{\bf u})$.
\end{enumerate}
\end{lemma}

{\bf Remark}.
The fact that the constant $C_0$ diverges as one approaches the apparent horizon is clear, since ${\bf X}$ becomes null when $r\to 2m$, see Lemma~\ref{Lem:XProperties}(b), and consequently ${\bf T}({\bf X},{\bf u})$ becomes degenerate on $r=2m$.\\

\proofof{Lemma~\ref{Lem:TXu}} Using Lemma~\ref{Lem:DEC}, we estimate
$$
{\bf T}({\bf X},{\bf Y}) = \sqrt{1 + 2E(R)} {\bf T}({\bf u},{\bf Y}) - \dot{r} {\bf T}({\bf w},{\bf Y})
 \geq \left( \sqrt{1 + 2E(R)} - |\dot{r}| \right) {\bf T}({\bf u},{\bf Y}).
$$
Since $1 + 2E(R) - |\dot{r}|^2 = 1 - 2m(R)/r$ we find
$$
{\bf T}({\bf u},{\bf Y}) \leq \frac{\sqrt{1 + 2E(R)} + |\dot{r}|}{1 - \frac{2m(R)}{r}}
{\bf T}({\bf X},{\bf Y})
$$
outside the apparent horizon. Since $E(R)\leq 0$ and $|\dot{r}| = \sqrt{2E(R) + 2m(R)/r} \leq  \chi$ we can further estimate
$$
\frac{\sqrt{1 + 2E(R)} + |\dot{r}|}{1 - \frac{2m(R)}{r}} \leq \frac{1 + \chi}{1 - \chi^2}
 = \frac{1}{1-\chi} \leq C_0,
$$
and inequality (a) follows. The inequalities (b) and (c) are a direct consequence of (a) combined with Lemma~\ref{Lem:DEC}. Finally, for inequality (d) we notice that
$$
T({\bf u},{\bf X}) = \gamma r' {\bf T}({\bf u},{\bf u}) - \dot{r}  {\bf T}({\bf u},{\bf w})
 \leq  (\sqrt{1 + 2E} + \chi) {\bf T}({\bf u},{\bf u}) \leq 2{\bf T}({\bf u},{\bf u}).
$$
\qed

\begin{figure}[h!]
\begin{center}
\includegraphics[width=9.cm]{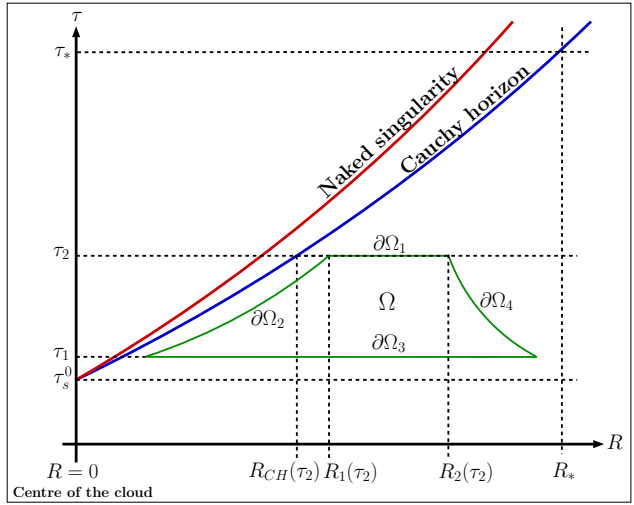}
\end{center}
\caption{\label{Fig:Estimate_region2} Illustration of the domain $\Omega$ and the functions involved in the estimate of the source term $S_X$ and the proof of integrability of the function $\alpha(\tau)$.}
\end{figure}

Using Eq.~(\ref{Eq:First_estimate}) and the result from Lemma~\ref{Lem:TXu} we find
\begin{equation}
\int_\Omega S_{\bf X} \sqrt{|g|} d^4x 
 \leq \int_{\tau_1}^{\tau_2} \alpha(\tau) {\cal E}_{\bf X}(\tau) d\tau,
\label{Eq:Second_estimate}
\end{equation}
with the nonnegative function $\alpha: [0,\infty)\to\Real$ defined by
$$
\alpha(\tau) := 4\pi C_0 
\sup_{R_1(\tau) \leq R \leq R_2(\tau)} \left[ (r\rho)(\tau,R) \right],\qquad
\tau\geq 0.
$$
It remains to show that $\alpha$ is integrable. To this purpose, we partition the time interval into the three intervals $[0,\tau_s^0)$, $(\tau_s^0,\tau_*)$ and $(\tau_*,\infty)$, where we recall that $\tau_s^0 := \tau_s(0)$ is the proper time at which the central naked singularity forms and $\tau_*$ is the proper time of a co-moving observer along the surface of the cloud at the moment the Cauchy horizon emerges from it, see figure~\ref{Fig:Estimate_region2}. In the following, we show that $\alpha$ is integrable on each of these intervals, implying the statement of Proposition~\ref{Prop:alpha}.

\noindent
\emph{First interval} ($0 \leq \tau < \tau_s^0$):

From Eq.~(\ref{Eq:SX}), it follows that there exists a constant $C_1 > 0$ such that
$$
r\rho \leq \frac{C_1}{y^{5/2}}
$$
in the region $D$. Using Lemma~\ref{Lem:Bounds}, we can make the constant larger if necessary, such that
\begin{equation}
r\rho \leq \frac{C_1}{(\tau_s(R) - \tau)^{5/6}}
\label{Eq:rrhoEstimate}
\end{equation}
in $D$. For $\tau < \tau_s^0$ we have $R_1(\tau) = 0$, and since $\tau_s(R) \geq \tau_s^0$ we find
$$
\alpha(\tau) \leq \sup\limits_{0\leq R \leq R_2(\tau)} \frac{4\pi C_0 C_1}{(\tau_s(R) - \tau)^{5/6}}
 \leq \frac{4\pi C_0 C_1}{(\tau_s^0 - \tau)^{5/6}},\qquad
0\leq \tau < \tau_s^0,
$$
which is integrable on the interval $[0,\tau_s^0)$.\\

\noindent
\emph{Second interval} ($\tau_s^0 < \tau < \tau_*$):

In this case we can still use the estimate~(\ref{Eq:rrhoEstimate}) and obtain the bound
$$
\alpha(\tau) \leq \sup\limits_{R_1(\tau)\leq R \leq R_2(\tau)} 
\frac{4\pi C_0 C_1}{(\tau_s(R) - \tau)^{5/6}} 
 = \frac{4\pi C_0 C_1}{(\tau_s(R_1(\tau)) - \tau)^{5/6}}
 \leq \frac{4\pi C_0 C_1}{(\tau_s(R_{CH}(\tau)) - \tau)^{5/6}},
$$
by the monotonicity of the function $\tau_s$, where $R_{CH}(\tau)$ is the value of $R$ at the intersection of the hypersurface $\Sigma_\tau$ with the Cauchy horizon, see figure~\ref{Fig:Estimate_region2}. The delicate part for the proof of the integrability of the function $\alpha$ relies on the following estimate whose proof is given in Appendix~\ref{App:Technical}.

\begin{lemma}
\label{Lem:Technical}
There are constants $\delta > 0$ and $K > 0$ such that
\begin{equation}
\tau_s(R_{CH}(\tau)) - \tau \geq K \left( \tau - \tau_s^0 \right)^{6/7}
\end{equation}
for all $\tau_s^0 < \tau < \tau_s^0 + \delta$.
\end{lemma}

Using this lemma we find that
$$
\alpha(\tau) \leq \frac{4\pi C_0 C_1}{K^{5/6}} \frac{1}{(\tau - \tau_s^0)^{5/7}}
$$
for all $\tau_s^0 < \tau < \tau_s^0 + \delta$, and hence $\alpha$ is integrable on the interval $(\tau_s^0,\tau_s^0 + \delta)$. Furthermore, $\alpha$ is also integrable on the remaining part $[\tau_s^0 + \delta,\tau_*)$ since there $\tau_s(R_{CH}(\tau)) - \tau$ is bounded away from zero.\\

\noindent
\emph{Third interval} ($\tau > \tau_*$):

In this case, $\rho=0$ and hence $\alpha(\tau) = 0$ for $\tau > \tau_*$ is trivially integrable.
\qed

\subsection{Main result}
\label{Sub:Main}

Combining the balance law equation from Proposition~\ref{Prop:Balance} with the result from Proposition~\ref{Prop:alpha} we arrive at the main conclusion of this section:

\begin{theorem}
\label{Thm:Main}
Consider a globally naked singularity in a TB spacetime $(M,{\bf g})$ satisfying the assumptions (i)-(viii) stated in section~\ref{Sec:Preliminaries}. Let $D \subset M$ be the maximal future development of the initial surface $\tau=0$ (see figure~\ref{Fig:D_epsilon}), and let ${\bf T}$ be a divergence-free stress-energy tensor satisfying the dominant energy condition on $M$. Then, there exists a constant $C > 0$ such that for every spacelike surface $\Sigma_\tau$ of constant $\tau > 0$ in $D$ and every null surface $\Sigma_U$ and $\Sigma_V$ of constant $U$ and $V$, respectively, which are contained in $D$, the following inequalities hold:
\begin{eqnarray}
\int_{\Sigma_\tau} \langle r^2 {\bf T}({\bf u},{\bf u}) \rangle_{S^2} \frac{dR}{\gamma} &\leq& C
 \qquad \mbox{(uniform energy bound)},
\label{Eq:Teo_eq1}\\
\int_{\Sigma_U} \langle r^2 {\bf T}({\bf u},{\bf k}) \rangle_{S^2} \frac{dV}{\dot{V}} &\leq& C 
\qquad \mbox{(uniformly bounded outgoing null flux)},
\label{Eq:Teo_eq2}\\
\int_{\Sigma_V} \langle r^2 {\bf T}({\bf u},{\bf l}) \rangle_{S^2} \frac{dU}{\dot{U}} &\leq& C 
\qquad \mbox{(uniformly bounded incoming null flux)}.
\label{Eq:Teo_eq3}
\end{eqnarray}
\end{theorem}

{\bf Remarks}:
\begin{enumerate}
\item Theorem~\ref{Thm:Main} can be applied to any test field $\Phi$ on $(M,{\bf g})$ whose dynamics is described by a diffeomorphism-invariant action and governed by a well-posed Cauchy problem and whose stress-energy tensor ${\bf T}$ satisfies the dominant energy condition, and thus it is rather general.
\item Physically, Theorem~\ref{Thm:Main} implies that an observer which is co-moving with the dust particles measures a finite energy of the test field $\Phi$ propagating on the TB background. In fact, the energy measured by the observer is uniformly bounded by the constant $C$, even as the observer crosses the Cauchy horizon at points lying arbitrarily close to the singularity.
\item The constant $C$ can be chosen proportional to the initial energy of the field: there is a constant $k > 0$ such that
\begin{equation}
C \leq k\int_{\Sigma_0} \langle r^2 {\bf T}({\bf u},{\bf u}) \rangle_{S^2} \frac{dR}{\gamma}.
\label{Eq:CEstimate}
\end{equation}
\end{enumerate}

\proofof{Theorem~\ref{Thm:Main}} Combining the results from Propositions~\ref{Prop:Balance} and~\ref{Prop:alpha} we find that
$$
{\cal E}_{\bf X}(\tau_2) - {\cal E}_{\bf X}(\tau_1) 
 \leq \int_{\tau_1}^{\tau_2}  \alpha(\tau) {\cal E}_{\bf X}(\tau) d\tau
$$
for all $\tau_2 > \tau_1 \geq 0$, where the function $\alpha: [0,\infty)\to \Real$ is nonnegative and integrable. Dividing both sides by $\tau_2 - \tau_1$ and taking the limit $\tau_2\to\tau_1$ gives
$$
\frac{d}{d\tau} {\cal E}_{\bf X}(\tau) \leq \alpha(\tau) {\cal E}_{\bf X}(\tau),\qquad
\tau \geq 0,
$$
from which it follows that
\begin{equation}
{\cal E}_{\bf X}(\tau) \leq {\cal E}_{\bf X}(0) e^{\int_0^\tau \alpha(s) ds} \leq C_1,
\qquad C_1 := {\cal E}_{\bf X}(0) e^{\int_0^\infty \alpha(s) ds} < \infty
\end{equation}
for all $\tau \geq 0$; thus it follows that ${\cal E}_{\bf X}$ is uniformly bounded by the constant $C_1$. Using this result in Proposition~\ref{Prop:Balance} and taking into account once again the integrability of the function $\alpha$, one obtains a bound for the flux integrals:
$$
{\cal F}_{\bf X}^-(U_2) + {\cal F}_{\bf X}^+(V_2)
 \leq {\cal E}_{\bf X}(\tau_1) 
 + \int_{\tau_1}^{\tau_2}  \alpha(\tau) {\cal E}_{\bf X}(\tau) d\tau
 \leq C_1 + C_1\int_0^\infty \alpha(\tau) d\tau =: C_2,
$$
which shows that the flux integrals ${\cal F}_{\bf X}^\pm$ are also uniformly bounded. Now the bounds (\ref{Eq:Teo_eq1},\ref{Eq:Teo_eq2},\ref{Eq:Teo_eq3}) follow using Lemma~\ref{Lem:TXu}(a). Finally, the estimate~(\ref{Eq:CEstimate}) follows from the definitions of $C_1$ and $C_2$ and Lemma~\ref{Lem:TXu}(d).
\qed

In the next section, we apply Theorem~\ref{Thm:Main} and certain generalizations of it to specific examples of physical interest.

\section{Applications}
\label{Sec:Applications}

In this section, we discuss a few applications of Theorem~\ref{Thm:Main}. We start by writing down explicitly the integrals which appear in the bounds in Theorem~\ref{Thm:Main} for the case of an electromagnetic test field and interprete them in physical terms. As a second application we consider a Klein-Gordon test field $\Phi$ and derive point-wise estimates for $\Phi$. These estimates will show that $\Phi$ cannot diverge along the Cauchy horizon (except, possibly, at the first singular point). Finally, we apply our results to odd-parity linearized gravitational perturbations of the collapsing dust spacetime and analyze the behaviour of the linearized Weyl scalar $\delta\Psi_0$ in the vicinity of the Cauchy horizon.

\subsection{Electromagnetic fields}
\label{Sub:Electromagnetic_fields}

For an electromagnetic test field propagating on $(M,{\bf g})$ the components of the stress-energy tensor are given by~\cite{Wald-Book}
$$
T^{\mu\nu} = F^{\mu\rho} F^\nu{}_\rho  - \frac{1}{4}g^{\mu\nu} F^{\sigma\rho} F_{\sigma\rho},
$$
where $F_{\mu \nu}$ are the components of the Faraday tensor. With respect to the local orthonormal frame $\{ {\bf u}, {\bf w}, {\bf e}_\vartheta, {\bf e}_\varphi \}$ the Faraday tensor has the usual components
$$
\left( F_{\alpha\beta} \right) 
 = \left( \begin{array}{cccc}
0      & -E_1 & -E_2 & -E_3\\
E_1 & 0       & B_3   & -B_2 \\
E_2 & -B_3 & 0        & B_1 \\
E_3 & B_2  & -B_1  & 0\\
\end{array} \right), \qquad
F^{\alpha \beta} F_{\alpha \beta} = -2\left( |\ve{E}|^2 - |\ve{B}|^2 \right)
$$
with $\ve{E} = (E_1,E_2,E_3)$ and $\ve{B} = (B_1,B_2,B_3)$ the spatial frame components of the electric and magnetic field, and $|\ve{E}|$ and $|\ve{B}|$ their magnitude. In terms of these fields, the integral quantities~(\ref{Eq:Teo_eq1},\ref{Eq:Teo_eq2},\ref{Eq:Teo_eq3}) in Theorem~\ref{Thm:Main} are
\begin{eqnarray}
&&\frac{1}{2} \int_{\Sigma_\tau} 
 r^2 \langle |\ve{E}|^2 + |\ve{B}|^2 \rangle_{S^2} \frac{dR}{\gamma},\\
&&\frac{1}{2} \int_{\Sigma_U} 
 r^2 \langle |\ve{E}|^2 + |\ve{B}|^2 - 2(E_2 B_3 - E_3 B_2) \rangle_{S^2} \frac{dV}{\dot{V}},\\
&&\frac{1}{2} \int_{\Sigma_V}
r^2 \langle |\ve{E}|^2 + |\ve{B}|^2 + 2(E_2 B_3 - E_3 B_2) \rangle_{S^2} \frac{dU}{\dot{U}},
\end{eqnarray}
respectively. The terms $|\ve{E}|^2 + |\ve{B}|^2$  and $E_2 B_3 - E_3 B_2$ correspond to the energy density and radial component of the Poynting vector associated with the electromagnetic field. The uniform bounds in Theorem~\ref{Thm:Main} thus imply that the total electromagnetic energy and radiation flux through the surfaces $\Sigma_U$ and $\Sigma_V$ are bounded as we approach the singularity.

\subsection{Linear scalar fields}
\label{Sub:Scalar_fields}

As a next example we consider a real-valued test field $\Phi$ of mass $\mu\geq 0$ satisfying the Klein-Gordon equation,
\begin{equation}
\Box \Phi + \mu^2\Phi = 0,
\label{Eq:KG}
\end{equation}
where $\Box := -\nabla^{\mu} \nabla_{\mu}$ denotes the d'Alembertian operator on the spacetime $(M, {\bf g})$. The components of the stress-energy tensor associated with $\Phi$ are given by~\cite{Wald-Book}
\begin{equation}
T_{\mu\nu} := \nabla_{\mu} \Phi \cdot \nabla_{\nu} \Phi
 - \frac{1}{2} g_{\mu\nu} \left( \nabla^\rho \Phi \cdot \nabla_\rho \Phi + \mu^2 \Phi^2 \right).
\label{Eq:TmunuScalarField}
\end{equation}
In terms of the orthonormal frame $\{ {\bf u}, {\bf w}, {\bf e}_\vartheta, {\bf e}_\varphi \}$, we find $\nabla_0\Phi = \dot{\Phi}$, $\nabla_1\Phi = \gamma\Phi'$, $\nabla_A\Phi = \hat{\nabla}_A\Phi/r$, with $A=\vartheta,\varphi$, and $\hat{\nabla}$ denoting the covariant derivative associated with the standard metric on the two-sphere, and thus the quantities~(\ref{Eq:Teo_eq1},\ref{Eq:Teo_eq2},\ref{Eq:Teo_eq3}) in Theorem~\ref{Thm:Main} yield, respectively,
\begin{eqnarray}
&&\frac{1}{2} \int_{\Sigma_\tau} r^2 \langle \dot{\Phi}^2 + (\gamma \Phi')^2 + \frac{1}{r^2}\hat{\nabla}^A \Phi \hat{\nabla}_A \Phi + \mu^2 \Phi^2 \rangle_{S^2} \frac{dR}{\gamma}, \label{Eq:Energy_estimate}\\
&&\frac{1}{2} \int_{\Sigma_U} r^2 \langle \left( \dot{\Phi} + \gamma\Phi' \right)^2 + \frac{1}{r^2}\hat{\nabla}^A \Phi \hat{\nabla}_A \Phi + \mu^2 \Phi^2 \rangle_{S^2} \frac{dV}{\dot{V}},\nonumber \\
&&\frac{1}{2} \int_{\Sigma_V} r^2 \langle \left( \dot{\Phi} - \gamma\Phi' \right)^2 + \frac{1}{r^2}\hat{\nabla}^A \Phi \hat{\nabla}_A \Phi + \mu^2 \Phi^2 \rangle_{S^2} \frac{dU}{\dot{U}}.\nonumber
\end{eqnarray}
The uniform boundedness of the integral~(\ref{Eq:Energy_estimate}) has the following important implication, which generalizes and strengthens the result of Theorem~2 in Ref.~\cite{nOoS14}:

\begin{theorem}
\label{Thm:KG}
Consider a solution $\Phi$ of the Klein-Gordon equation~(\ref{Eq:KG}) on the spacetime manifold $(D,{\bf g})$ belonging to initial data for $\Phi$ and $\dot{\Phi}$ on the Cauchy surface $\tau = 0$ which is smooth and has compact support. 

Then, there exists a positive constant $C$ such that for all $x\in D$
\begin{equation}
\sqrt{r}|\Phi(x)| \leq C.
\label{Eq:KGBounds}
\end{equation}
\end{theorem}

{\bf Remarks}:
\begin{enumerate}
\item In Theorem~2 of~\cite{nOoS14} we had proven that a rotationally symmetric field $\Phi$ satisfying the Klein-Gordon equation satisfies the bound $r|\Phi(x)| \leq C_1$ for some constant $C_1$. Since our new bound~(\ref{Eq:KGBounds}) involves the square root of $r$ instead of $r$, it constitutes an improved bound on the behaviour of $\Phi$ near the central singular point.
\item Although the bound in Eq.~(\ref{Eq:KGBounds}) does not exclude the possibility that $\Phi$ diverges at the first singular point, it does rule out that such a hypothetical divergence would propagate along the Cauchy horizon. Thus, the importance of this bound relies in the fact that it proves that the field $\Phi$ is uniformly bounded away from the central singularity. In fact, Eq.~(\ref{Eq:KGBounds}) also implies that $\Phi$ decays along the Cauchy horizon as $r\to \infty$. However, once we know that $\Phi$ is bounded at the surface of the cloud, this result is a direct consequence of the known stability results for the wave equation on a Schwarzschild background, see for example~\cite{bKrW87,mDiR08}.
\item Numerical investigations performed by one of us~\cite{Nestor-PhD-thesis} suggest that, in fact, the field $\Phi$ is bounded everywhere on the Cauchy horizon (including the vicinity of the central singularity), so likewise, the bound~(\ref{Eq:KGBounds}) is not optimal and could be further improved.
\end{enumerate}

\proofof{Theorem~\ref{Thm:KG}} By standard theorems on existence and uniqueness of solutions of the wave equation on globally hyperbolic spacetimes (see, for instance Theorem~10.1.2 in~\cite{Wald-Book}) the solution $\Phi$ is smooth and vanishes outside a large enough sphere on each surface $\Sigma_\tau$ of constant $\tau$. Therefore, we can decompose $\Phi$ in terms of standard spherical harmonics $Y^{\ell m}$:
\begin{equation*}
\Phi = \sum_{\ell m} \phi_{\ell m}Y^{\ell m},
\end{equation*}
with smooth functions $\phi_{\ell m}$ of $(\tau,r)$ which, for each $\tau\geq 0$, vanish for $r$ large enough. Using integration by parts and the fact that the standard spherical harmonics provide an orthonormal basis of $L^2(S^2)$ which diagonalize the Laplacian on $S^2$, we have
\begin{eqnarray*}
\langle \Phi^2 \rangle_{S^2} &=& \sum\limits_{\ell m} |\phi_{\ell m}|^2,\\
\langle \hat{\nabla}^A \Phi \hat{\nabla}_A \Phi \rangle_{S^2} 
 &=& \sum\limits_{\ell m} \ell(\ell+1) |\phi_{\ell m}|^2.
\end{eqnarray*}
In terms of the {\it generalized angular derivative operator} (see Appendix~\ref{App:Sobolev})
$$
D^{\sigma}\Phi := \sum_{\ell m}(2\ell + 1)^{\sigma} \phi_{\ell m}Y^{\ell m}
$$
we thus have
$$
\langle \hat{\nabla}^A \Phi \hat{\nabla}_A \Phi \rangle_{S^2} 
 \geq \frac{1}{5}\langle (D^1\Phi_{\ell>0})^2 \rangle_{S^2},
$$
where we have used the inequality $(2\ell + 1)^2 \leq 5\ell (\ell + 1)$ which is valid for all $\ell \geq 1$, and where here and in the following $\Phi_{\ell>0} := \Phi - \langle \Phi \rangle_{S^2}/(4\pi)$ refers to the non-spherical part of $\Phi$. Using these preliminary remarks, the bound $0 < \delta \leq \gamma r' \leq 1$, and the fact that along $\Sigma_\tau$ we have $dr = r' dR$, the energy bound in Eq.~(\ref{Eq:Energy_estimate}) implies that
\begin{equation*}
\int_{r_1(\tau)}^{\infty} 
\langle r^2\dot{\Phi}^2 + r^2\Phi_r^2 +  (D^1\Phi_{\ell>0})^2 
 + \mu^2 r^2\Phi^2\rangle_{S^2} dr
 \leq C,
\end{equation*}
with $C$ independent of $\tau$ and $r_2$. Here, $\Phi_r$ denotes the partial derivative of $\Phi$ with respect to $r$ at fixed $\tau$, and $r_1(\tau)$ denotes the minimum of $r$ along $\Sigma_\tau$, that is, $r_1(\tau) = 0$ if $\tau < \tau_s^0$ and $r_1(\tau) = r_{CH}(\tau)$ otherwise. Because of the spherical symmetry of the background, the fields $D^\sigma \Phi$ also satisfy the Klein-Gordon equation~(\ref{Eq:KG}) with smooth initial data, and thus we obtain similar bounds when $\Phi$ is replaced with $D^\sigma\Phi$:
\begin{equation}
\int_{r_1(\tau)}^\infty 
\langle r^2\left| D^\sigma\dot{\Phi} \right|^2 + r^2\left| D^{\sigma}\Phi_r \right|^2 
 +  \left| D^{\sigma+1}\Phi_{\ell>0} \right|^2 
 + \mu^2 r^2\left| D^\sigma\Phi \right|^2 \rangle_{S^2} dr \leq C_\sigma,
\label{Eq:EnergyEstimateKG}
\end{equation}
for all $\sigma\geq 0$. Now the bound in Eq.~(\ref{Eq:KGBounds}) is a direct consequence of the following Sobolev-type estimate whose proof is provided in Appendix~\ref{App:Sobolev}:

\begin{lemma}[Sobolev estimate]
\label{Lem:Sobolev}
Let $r_1\geq 0$, $\Sigma := (r_1,\infty) \times S^2$. For each $\sigma > 1/2$ there exists a constant $K_\sigma > 0$ such that for all $\Phi\in C^\infty(\Sigma)$ which vanish for $r$ large enough, the following inequality holds:
\begin{equation*}
|\Phi(x)|^2 \leq \frac{K_\sigma}{r}
\int_{r_1}^{\infty} \langle r^2|D^{\sigma} \Phi_r |^2 
 + |D^{1+\sigma} \Phi_{\ell>0}|^2 \rangle_{S^2} dr,\qquad
x = (r,\hat{x})\in \Sigma.
\end{equation*}
\end{lemma}

Note that it is not possible to improve much the point-wise estimate~(\ref{Eq:KGBounds}) based solely on the estimate~(\ref{Eq:EnergyEstimateKG}). For example, the time-independent, spherically symmetric function $\Phi = r^{-\alpha}$ with $0 < \alpha < 1/2$ satisfies
$$
r^2\Phi_r^2 + \mu^2 r^2 \Phi^2 = (\alpha^2 + \mu^2 r^2) r^{-2\alpha},
$$
which is integrable near $r = 0$. Multiplying this function with a smooth cutoff function we obtain a new function $\Phi$ which satisfies the bound~(\ref{Eq:EnergyEstimateKG}) but diverges at $r=0$ with a power of $1/r$ lying arbitrarily close to $1/2$.
\qed

\subsection{Odd-parity linearized gravitational and dust perturbations}
\label{Sub:GravPert}

As a final application of Theorem~\ref{Thm:Main} we consider gravitational linear perturbations of the dust collapse model described in section~\ref{Sec:Preliminaries}. For simplicity, we focus our attention on perturbations with odd parity, in which case the dynamics of the perturbations can be described by a single master equation for a gauge-invariant linear combination of the perturbed metric fields with a source term that depends on the vorticity perturbations of the dust field. To our knowledge, this problem has first been addressed by Harada {\it et al} \cite{hItHkN98,hItHkN99} in the marginally bound case based on numerical simulations. When the source term is zero they show that the linear quadrupolar metric perturbations remain finite at the Cauchy horizon. However, their numerical results also show that a nontrivial vorticity perturbation of the dust distribution induces a divergence of the gravitational field at the central singularity, characterized by a power-law like behaviour of the linearized Weyl scalar $\delta\Psi_0$ along the centre which is of the form~\cite{hItHkN99}
\begin{equation}
\im(\delta\Psi_0) \sim \left( \tau_s^0 - \tau \right)^{-5/3},\qquad \tau\to \tau_s^0,
\label{Eq:Psi0PowerLaw}
\end{equation}
though they also find that this divergence does not propagate along the Cauchy horizon and, in this sense, cannot act as a strong source of gravitational waves. A more recent line of work based on analytic methods was initiated by Duffy and Nolan~\cite{eDbN11b} who analyzed the linear metric perturbations of self-similar, marginally bound collapse and proved that in the odd-parity sector these remain bounded at the Cauchy horizon.

The goal of this section is to derive bounds for the odd-parity linearized perturbations in the vicinity of the Cauchy horizon, extending the results of~\cite{hItHkN98,hItHkN99} to the more generic bounded collapse case and putting them on a rigorous basis.

\subsubsection{Perturbation equations}

The equations governing the dynamics of linearized odd-party perturbations of a self-gravitating spherical perfect fluid configuration can be represented in the following form (see Eq. (96) in Ref.~\cite{eCnOoS13} and references therein for further details)
\begin{eqnarray}
\tilde{*} d(r^2{\cal F}) - (\hat{\Delta} + 2){\bf h} &=& 16\pi r^2 n\omega\underline{\bf u},
\label{Eq:OddParityEq1}\\
\tilde{d}^\dagger {\bf h} &=& 0.
\label{Eq:OddParityEq2}
\end{eqnarray}
Here, the operators $d$, $\tilde{*}$ and $\tilde{d}^\dagger = \tilde{*} d\tilde{*}$ denote the exterior differential, the Hodge dual and the codifferential on the two-dimensional Lorentzian manifold $\tilde{M}$ with metric $\tilde{\bf g} = -d\tau^2 + \gamma^{-2} dR^2$ corresponding to the radial part of the background metric~(\ref{Eq:MetricSol}), and $\hat{\Delta}$ refers to the Laplacian on the two-sphere. The one-form ${\bf h}$ refers to the Gerlach-Sengupta gauge-invariant combination~\cite{uGuS79} of the odd-parity metric perturbations, and ${\cal F} := r^2\tilde{*} d(r^{-2}{\bf h})$ is a gauge-invariant scalar field which is related to the imaginary part of the linearized Weyl scalar $\delta\Psi_0$ as follows~\cite{rP72b}:
\begin{equation}
\im\left( \delta \Psi_0 \right) 
 = \im\left[ \delta R_{\alpha\beta\gamma\delta} k^\alpha m^\beta(m^*)^\gamma l^\delta \right]
 = \frac{\hat{\Delta}{\cal F}}{r^2}.
\label{Eq:deltaPsi0}
\end{equation}
Furthermore, $n$ and $\underline{\bf u} = u_\mu dx^\mu$ refer to the particle density and the one-form associated with the four-velocity of the background fluid, and $\omega$ parametrizes the vorticity perturbations of the fluid (see~\cite{eCnOoS13}). The linearized Euler equations imply that $\pounds_{\bf u}\omega = 0$, such that $\omega$ is constant along the flow lines. The regularity conditions imply that $\omega = R v$ with a smooth function $v\in C^\infty(\Sigma_0)$ on the initial time slice which can be decomposed into its dipole and dipole-free parts:
$$
v = v_{\ell=1} + v_{\ell > 1}
$$
(the monopole part is absent because $\omega$ is only defined up to a constant). By regularity, $v_{\ell=1} = {\cal O}(R)$ and $v_{\ell > 1} = {\cal O}(R^2)$. Without loss of generality we assume that $v = 0$ for $R > R_*$ outside the collapsing cloud.

In our case of a dust fluid, we have simply $\underline{\bf u} = -d\tau$ and $n$ is proportional to the mass density $\rho$, so we can replace $n$ with $\rho$ and absorb the constant into $\omega$. Following the procedure outlined in~\cite{eCnOoS13}, one derives the following master equation from Eqs.~(\ref{Eq:OddParityEq1},\ref{Eq:OddParityEq2}):
\begin{equation}
\tilde{\Box}\psi + V_{eff}\psi = -16\pi \gamma r\frac{\partial}{\partial R}(\rho\omega),
\label{Eq:OddParityMasterEq1}
\end{equation}
for the scalar quantity $\psi := r{\cal F}$, where $\tilde{\Box} := \tilde{d}^\dagger d = -\tilde{g}^{ab}\tilde{\nabla}_a\tilde{\nabla}_b$ denotes the covariant d'Alembert operator on $(\tilde{M},\tilde{\bf g})$ and
\begin{equation}
V_{eff} = -\frac{1}{r^2}\hat{\Delta} - \frac{6m}{r^3} + 4\pi \rho.
\label{Eq:Veff}
\end{equation}
This equation is equivalent to the one used for the numerical simulations presented in~\cite{hItHkN98,hItHkN99} and yields a master equation for $\im(\delta\Psi_0) = \hat{\Delta}\psi/r^3$.

However, for our results below, a different master equation will turn out to be more useful which is obtained as follows: first, one uses Eq.~(\ref{Eq:OddParityEq2}) in order to introduce a potential function $\phi$ such that ${\bf h} = \tilde{*} d(r\phi)$. Eq.~(\ref{Eq:OddParityEq1}) can then be integrated and yields (up to a constant that can be reabsorbed in the definition of $\phi$)
\begin{equation}
r{\cal F} - (\hat{\Delta} + 2)\phi = r^3 S_{dust},
\label{Eq:OddParityEq3}
\end{equation}
with
\begin{equation}
S_{dust} 
 = \frac{16\pi}{r^4}\int_0^R 
 r^2\rho(y,\bar{R}) \omega(\bar{R},\vartheta,\varphi) \frac{d\bar{R}}{\gamma}
 = \frac{16\pi}{r^4}\int_0^R \frac{\bar{R}^3\rho_0(\bar{R}) v(\bar{R},\vartheta,\varphi)}
 {\sqrt{1 + 2E(\bar{R})}} d\bar{R}.
\label{Eq:Sdust}
\end{equation}
Reexpressing ${\cal F}$ in Eq.~(\ref{Eq:OddParityEq3}) in terms of $\phi$ one obtains the new master equation
\begin{equation}
\tilde{\Box} \phi + V_{eff}\phi = r S_{dust},
\label{Eq:OddParityMasterEq2}
\end{equation}
with the same effective potential $V_{eff}$ as the one appearing in Eq.~(\ref{Eq:OddParityMasterEq1}), but with a different source term involving an integral of $\rho\omega$ instead of its derivative. Finally, by setting $\Phi = \phi/r$, one can rewrite the master equation~(\ref{Eq:OddParityMasterEq2}) in its $3+1$ spacetime version [instead of the $1+1$ form on $(\tilde{M},\tilde{\bf g})$] which will be more useful for the applications of our results:
\begin{equation}
\Box \Phi + V\Phi = S_{dust},\qquad
V = -\frac{8m}{r^3} + 8\pi \rho.
\label{Eq:KG_like}
\end{equation}
Once $\Phi$ has been determined, the imaginary part of the linearized Weyl scalar $\delta\Psi_0$ is obtained from Eqs.~(\ref{Eq:deltaPsi0}) and (\ref{Eq:OddParityEq3}) which yield
\begin{equation}
\im(\delta\Psi_0) = \hat{\Delta}\left[ S_{dust} + \frac{1}{r^2}(\hat{\Delta}+2)\Phi \right].
\label{Eq:ImdeltaPsi0}
\end{equation}

After these remarks about the perturbation equations and their relation to the linearized Weyl tensor, we formulate the main result of this section, which puts bounds on the behaviour of the dipole-free metric perturbations.

\subsubsection{Main bounds on the metric perturbations and the linearized Weyl scalar}

\begin{theorem}
\label{Thm:OddParity}
Consider a dipole-free solution $\Phi$ of the master equation~(\ref{Eq:KG_like}) on the spacetime manifold $(D,{\bf g})$ belonging to initial data for $\Phi$ and $\dot{\Phi}$ on the Cauchy surface $\tau = 0$ which is smooth and has compact support. Suppose the dipole-free contributions from the vorticity perturbations $\omega_{\ell > 1}$ satisfy the previously mentioned regularity conditions, that is $\omega_{\ell > 1}\in C^\infty(\Sigma_0)$, $\omega_{\ell > 1} = {\cal O}(R^3)$, and $\omega_{\ell > 1} = 0$ for $R > R_*$.

Then, there are positive constants $C_1$ and $C_2$ such that
\begin{equation}
\sqrt{r}|\Phi(x)| \leq C_1,\qquad
r^{5/2}|\im(\delta\Psi_0)| \leq C_2,
\label{Eq:GravBounds}
\end{equation}
for all $x\in D_c := \{ p\in D : R \leq R_* \}$ in the interior of the cloud.
\end{theorem}

{\bf Remarks}:
\begin{enumerate}
\item As for the case of the Klein-Gordon equation, our bounds~(\ref{Eq:GravBounds}) do not exclude the possibility that $\Phi$ or $\delta\Psi_0$ diverge at the central singularity; nevertheless they prevent the linearized gravitational field from becoming infinite along the Cauchy horizon away from the singularity.
\item Under the hypothesis of Theorem~\ref{Thm:OddParity}, the dipole-free contributions to the source term $S_{dust}$ are of the order of $R^6/r^4 = R^2/y^8 = (R^2/y^3) y^{-5}$, which in view of Eq.~(\ref{Eq:ImdeltaPsi0}) gives a contribution to $\im(\delta\Psi_0)$ of the same form. This contribution vanishes at the centre of the cloud; however it diverges as $1/y^5$ along the Cauchy horizon, which is a little bit slower than the divergence $1/y^6$ of the background solution [see Eq.~(\ref{Eq:Psi0Bound})]. Nevertheless, in our bound~(\ref{Eq:GravBounds}), this contribution could, in principle, be overshadowed by a more rapidly growing term of the order $r^{-5/2} = R^{-5/2} y^{-5}$ originating from the quantity $\Phi/r^2$ in the second term on the right-hans side of Eq.~(\ref{Eq:ImdeltaPsi0}). It would be desirable to have higher-order estimates for $\Phi$ in order to get rid of the factor $R^{-5/2}$ which diverges at the centre, and to compare the resulting improved estimates to the numerical empirical result~(\ref{Eq:Psi0PowerLaw}).
\item It should be possible to extend the bounds~(\ref{Eq:GravBounds}) to the exterior of the cloud, based on the known linear stability results for the Schwarzschild spacetime~\cite{bKrW87,mDiR08,mDgHiR16}. Since the main goal of this work is to understand the behaviour of the fields in the vicinity of the central singularity, we shall not pursue this issue here.
\item For dipole perturbations we have $(\hat{\Delta}+2)\Phi_{\ell=1} = 0$, and it follows from Eq.~(\ref{Eq:OddParityEq3}) that the vorticity induces the angular momentum (see, for instance Eq.~(10) in~\cite{oSmT01})
$$
J(R) := -\frac{1}{6} r^2{\cal F}_{10} = -\frac{8\pi}{3}\int_0^R \frac{\bar{R}^3\rho_0(\bar{R}) v_{10}(\bar{R})} {\sqrt{1 + 2E(\bar{R})}} d\bar{R} = {\cal O}(R^5),
$$
where we have expanded $v_{\ell=1}(R,\vartheta,\varphi) = v_{10}(R)\cos(\vartheta)$ and likewise for ${\cal F}$. Notice that for $R > R_*$ outside the cloud the quantity $J(R)$ is constant and describes the linearized Kerr mode~\cite{oSmT01}. Comparing the corresponding contribution to the linearized Weyl scalar $\delta\Psi_0$ with the background Weyl scalar $\Psi_0$ we obtain
$$
\frac{\im(\delta\Psi_0)_{10}}{\Psi_0} = \frac{12J(R)}{r^4\Psi_0}
 = \frac{12J(R)}{R^5}\frac{1}{y^6\Psi_0}\frac{R}{y^2}.
$$
In view of the limit in Eq.~(\ref{Eq:Psi0Bound}) the right-hand side diverges as the central singularity is approached along the Cauchy horizon, and in this sense vorticity dipole perturbations of the cloud yield large metric dipole perturbations in the vicinity of the central singularity, indicating that nonlinear terms are likely to become important.
\end{enumerate}

The proof of Theorem~\ref{Thm:OddParity} is based on the following theorem which generalizes the result of Theorem~\ref{Thm:KG} to the case of a wave equation with effective potential and source term:

\begin{theorem}
\label{Thm:KGBis}
Let $V,F: D\to \Real$ be real-valued $C^\infty$-differentiable functions satisfying the following assumptions:
\begin{enumerate}
\item[(a)] $V$ is rotational invariant and there exists a constant $K > 0$ such that
\begin{equation}
|V(x)| \leq K\frac{m(R)}{r^3}
\end{equation}
for all $x\in D$.
\item[(b)] The function $F$ satisfies
\begin{equation}
\int_{D(\varepsilon)} (D^\sigma F)^2 \sqrt{|g|} d^4x < \infty
\end{equation}
for some $\varepsilon > 0$ and some $\sigma > 1/2$.
\end{enumerate}
Let $\Phi$ be a monopole-free solution of the wave equation
\begin{equation}
\Box\Phi + V\Phi = F,
\end{equation}
on the spacetime manifold $(D,g)$ belonging to initial data $\Phi$ and $\dot{\Phi}$ on the Cauchy surface $\tau = 0$ which is smooth and has compact support.

Then, $\Phi$ satisfies the same uniform bound as in section~\ref{Sub:Scalar_fields}. In particular, there exists a positive constant $C$ such that
$$
\sqrt{r}|\Phi(x)| \leq C
$$
for all $x\in D_c$ in the interior of the cloud.
\end{theorem}

We postpone the proof of this theorem to the end of this section. We first show that it implies Theorem~\ref{Thm:OddParity}:\\

\proofof{Theorem~\ref{Thm:OddParity}}
In view of Theorem~\ref{Thm:KGBis}, it is sufficient to show that the functions $V$ and $S_{dust}$ defined in Eqs.~(\ref{Eq:KG_like}) and (\ref{Eq:Sdust}) satisfy assumptions (a) and (b), respectively. The statement of the theorem then follows from Eq.~(\ref{Eq:ImdeltaPsi0}) and the already established bounds on $S_{dust}$.

Regarding the condition on $V$ we first note that it is rotational invariant by construction. Next, using the explicit representation of $\rho$ [see Eqs.~(\ref{Eq:rho}) and (\ref{Eq:rprime})] we obtain
$$
V = -\frac{8m(R)}{r^3} 
 + \frac{8\pi\rho_0(R)}{y^6\left( 1 + \frac{R^2}{y^3} \sqrt{1 - q^2y^2}\Lambda(y,R) \right)}.
$$
The first term on the right-hand side obviously satisfies the required bound. The second term is bounded by a constant times $y^{-6} = R^3/r^3$ in $D(\varepsilon)$. Since $m(R) = {\cal O}(R^3)$ near the centre, and since $\rho_0(R) = 0$ for large $R$ the required bound for $V$ follows.

As for the source term $S_{dust}$ we have already established that in the dipole-free sector it can be bounded by a constant times $y^{-5}$ outside the Cauchy horizon. Using the first bound in Lemma~\ref{Lem:Estimates} we find
\begin{eqnarray*}
\int_{D(\varepsilon)} (D^\sigma S_{dust})^2 \sqrt{|g|} d^4x
 &=& \sum\limits_{\ell > 1}\sum\limits_{m=-\ell}^\ell (2\ell+1)^{2\sigma}
 \int_{\tau_1}^{\tau_2} \int_{R_1(\tau)}^{R_2(\tau)} 
 r^2 |S_{dust,\ell m}|^2 \frac{dR d\tau}{\gamma}\\
&\leq& const.\times\sum\limits_{\ell > 1}\sum\limits_{m=-\ell}^\ell (2\ell+1)^{2\sigma}
 \sup\limits_{R\geq 0} \left| \frac{v_{\ell m}(R)}{R^2} \right|^2
\int_{\tau_1}^{\tau_2} \int_{R_1(\tau)}^{R_2(\tau)} \frac{R^2}{y^4} dR d\tau.
\end{eqnarray*}
However, due to the regularity assumptions on the function $v = \omega/R$ and because
$$
\int_{R_1(\tau)}^{R_2(\tau)} \frac{R^2}{y^4} dR 
 = \int_{R_1(\tau)}^{R_2(\tau)} \left( \frac{R^2}{y^3} \right)^{4/3} R^{-2/3} dR
\leq const.\times  \int_{R_1(\tau)}^{R_2(\tau)} R^{-2/3} dR
\leq const.\times R_2(\tau)^{1/3},
$$
it follows that the integral of $(D^\sigma S_{dust})^2$ over $D(\varepsilon)$ is finite for all $\sigma\geq 0$, and Theorem~\ref{Thm:OddParity} is proven.
\qed

\subsubsection{Proof of Theorem~\ref{Thm:KGBis}}

It remains to show Theorem~\ref{Thm:KGBis}. Its proof proceeds along the same lines as the proof of Theorem~\ref{Thm:Main} and is based on the same stress-energy tensor as in the Klein-Gordon case, that is
\begin{equation}
T_{\mu\nu} = (\nabla_{\mu} \Phi)(\nabla_{\nu}\Phi)
 - \frac{1}{2} g_{\mu\nu} \left[ (\nabla^\rho\Phi)(\nabla_\rho\Phi) + \mu^2 \Phi^2 \right],
\end{equation}
for some positive constant $\mu > 0$. Due to the presence of the potential $V$ and the source term $F$, this stress-energy tensor is not divergence-free; it satisfies
$$
\nabla^{\mu}T_{\mu \nu} = -(\Box \Phi + \mu^2\Phi)\nabla_{\nu}\Phi 
 = -\left( \mu^2\Phi - V\Phi + F \right)\nabla_{\nu}\Phi.
$$
Therefore, the divergence of the current density  $J_{\bf X}^\mu = -T^\mu{}_\nu X^\nu$ is
\begin{equation}
\nabla_{\mu} J^{\mu}_{\bf X} = S_{\bf X} + S_{\bf X}^*,
\end{equation}
with the usual source term $S_{\bf X} = - T^{\mu \nu} \nabla_{\mu}X_{\nu}$ as defined in Eq.~(\ref{Eq:DivergenceLaw}) and the new source term
\begin{equation}
S_{\bf X}^* := -X_\nu\nabla_{\mu}T^{\mu \nu} = (\mu^2 - V)\Phi\nabla_{\bf X}\Phi 
 + F\nabla_{\bf X}\Phi,
\end{equation}
whose presence is due to the fact that $T_{\mu\nu}$ is not divergence-free anymore. Instead of the balance law in Proposition~\ref{Prop:Balance}, one obtains the new balance equation
\begin{equation}
{\cal E}_{\bf X}(\tau_2) + {\cal F}_{\bf X}^-(U_2) + {\cal F}_{\bf X}^+(V_2) 
 = {\cal E}_{\bf X}(\tau_1) + \int_\Omega (S_{\bf X} + S_{\bf X}^*) \sqrt{|g|} d^4x,
\label{Eq:EnergyBalanceBis}
\end{equation}
and it remains to bound the source terms $S_{\bf X}$ and $S_{\bf X}^*$.

For the following, we choose the timelike vector field ${\bf X}$ as in Eq.~(\ref{Eq:X_def}). The term involving $S_{\bf X}$ can then be estimated exactly as in Proposition~\ref{Prop:alpha}, such that
$$
\int_\Omega S_{\bf X} \sqrt{|g|} d^4x \leq 
\int_{\tau_1}^{\tau_2} \alpha(\tau) {\cal E}_{\bf X}(\tau) d\tau,
$$
with a nonnegative integrable function $\alpha: [0,\infty)\to\Real$. In order to bound the term involving the source term $S_{\bf X}^*$ we first observe that $|\dot{r}| = \sqrt{2E + 2m/r}\leq \sqrt{2m/r}\leq 1$, implying
$$
|\nabla_{\bf X}\Phi|^2 = |\sqrt{1+2E}\nabla_{\bf u}\Phi - \dot{r}\nabla_{\bf w}\Phi|^2
 \leq 2|\nabla_{\bf u}\Phi|^2 + 2|\nabla_{\bf w}\Phi|^2.
$$
Consequently,
\begin{eqnarray*}
\int_\Omega \mu^2\Phi\nabla_{\bf X}\Phi \sqrt{|g|} d^4x
 &\leq& \frac{\mu}{2}\int_\Omega \left( |\nabla_{\bf X}\Phi|^2 + \mu^2|\Phi|^2 \right)
  \sqrt{|g|} d^4x
\\
 &\leq& \mu\int_\Omega 
 \left( |\nabla_{\bf u}\Phi|^2 + |\nabla_{\bf w}\Phi|^2 + \frac{\mu^2}{2}|\Phi|^2 \right)
\sqrt{|g|} d^4x
\\
 &\leq& 2\mu\int_\Omega {\bf T}({\bf u},{\bf u}) \sqrt{|g|} d^4x
\\
 &\leq& 2\mu C_0 \int_{\tau_1}^{\tau_2} {\cal E}_{\bf X}(\tau) d\tau,
\end{eqnarray*}
where we have used Lemma~\ref{Lem:TXu} in the last step. Next, we estimate the second term in $S_{\bf X}^*$ as follows:
\begin{eqnarray*}
-\int_\Omega V\Phi\nabla_{\bf X}\Phi \sqrt{|g|} d^4x
 &\leq& \frac{1}{2}\int_\Omega |rV|
 \left( |\nabla_{\bf X}\Phi|^2 + \frac{|\Phi|^2}{r^2} \right) \sqrt{|g|} d^4x
\\
 &\leq& \frac{1}{2}\int_{\tau_1}^{\tau_2} \sup\limits_{R_1(\tau)\leq R \leq R_2(\tau)}
\left[ |rV|(\tau,R) \right] \left[ \int_{\Sigma_\tau\cap\Omega}
 \left( |\nabla_{\bf X}\Phi|^2 + \frac{|\Phi|^2}{r^2} \right) \right] d\tau.
\end{eqnarray*}
Since we are only considering multipoles $\ell\geq 1$ in $\Phi$, the expression inside the square parenthesis can be bounded from above by $4C_0{\cal E}_{\bf X}(\tau)$, as before. Therefore, we obtain
$$
-\int_\Omega V\Phi\nabla_{\bf X}\Phi \sqrt{|g|} d^4x \leq
\int_{\tau_1}^{\tau_2} \alpha^*(\tau) {\cal E}_{\bf X}(\tau) d\tau.
$$
with
$$
\alpha^*(\tau) := 2C_0\sup\limits_{R_1(\tau)\leq R \leq R_2(\tau)} \left[ |rV|(\tau,R) \right],
\qquad \tau\geq 0.
$$
According to assumption (a), one has $|r V|\leq K m(R)/r^2$ which is bounded outside the cloud because $m(R)$ is constant there. In the interior of the cloud one has instead the estimate
$$
2|rV| \leq K\frac{R c(R)}{y^4} \leq K c(0)\left( \frac{R^2}{y^3} \right)^{1/2}\frac{1}{y^{5/2}},
$$
which can be bounded by a constant $C_1$ divided by $y^{5/2}$. Therefore, we have exactly the same estimate as in Eq.~(\ref{Eq:Second_estimate}) with $r\rho$ replaced with $2|r V|$ which can both be bounded from above by $C_1/y^{5/2}$. From the proof of Proposition~\ref{Prop:alpha} it follows that $\alpha^*: [0,\infty)\to \Real$ is a nonnegative integrable function.

Finally, the third term in $S_{\bf X}^*$ can be estimated according to
$$
\int_\Omega F\nabla_{\bf X}\Phi \sqrt{|g|} d^4x
\leq \frac{1}{2}\int_\Omega\left( |\nabla_{\bf X}\Phi|^2 + F^2 \right) \sqrt{|g|} d^4x
\leq \frac{1}{2}\int_{\tau_1}^{\tau_2} \left( 4C_0{\cal E}_{\bf X}(\tau) 
 + \int_{\Sigma_\tau\cap\Omega} F^2\right) d\tau.
$$

Gathering the results, the balance law equation~(\ref{Eq:EnergyBalanceBis}) implies the following energy estimate:
$$
{\cal E}_{\bf X}(\tau_2) - {\cal E}_{\bf X}(\tau_1)
 \leq \int_{\tau_1}^{\tau_2} \left( \tilde{\alpha}(\tau) {\cal E}_X(\tau) 
 + \frac{1}{2}\int_{\Sigma_\tau\cap\Omega} F^2\right) d\tau,\qquad
\tau_2 > \tau_1 > 0,
$$
with the nonnegative function $\tilde{\alpha} := \alpha + \alpha^* + 2(1+\mu)C_0 : [0,\infty)\to \Real$. Due to the presence of the positive constants $2(1+\mu)C_0$ this function is not integrable on $[0,\infty)$; however since we are only interested in obtaining a uniform bound in the interior of the cloud the only relevant feature of $\tilde{\alpha}$ needed in our proof is its integrability on the interval $[0,\tau_*]$. Dividing both sides of the inequality by $\tau_2 - \tau_1 > 0$ and taking the limit $\tau_2\to\tau_1$ yields
$$
\frac{d}{d\tau} {\cal E}_{\bf X}(\tau) \leq \tilde{\alpha}(\tau){\cal E}_{\bf X}(\tau)
 + \frac{1}{2}\int_{\Sigma_\tau\cap\Omega} F^2,
$$
from which we obtain the required bound
$$
{\cal E}_{\bf X}(\tau) \leq e^{\int_0^\tau \tilde{\alpha}(s) ds}
\left( {\cal E}_{\bf X}(0) + \frac{1}{2}\int_D F^2  \sqrt{|g|} d^4x \right)
$$
for ${\cal E}_X(\tau)$ which yields a uniform bound for ${\cal E}_X$ inside the cloud. Going back to Eq.~(\ref{Eq:EnergyBalanceBis}) one obtains similar uniform bounds for the fluxes. Finally, the point-wise bound on $\Phi$ is obtained in exactly the same fashion as in section~\ref{Sub:Scalar_fields}.
\qed

\section{Conclusion}
\label{Sec:Discussion}

Within the framework of the Tolman-Bondi (TB) spacetime, which provides a relativistic model describing the gravitational collapse of a finite spherical dust cloud, globally naked singularities are known to appear under generic initial conditions consisting of inhomogeneous, spherical dust distributions. Initial data causing the formation of a shell-focusing curvature singularity leads to a spacetime possessing a Cauchy horizon which is generated by the first outgoing null geodesic emanating from the singularity. When the Cauchy horizon extends all the way to future null infinity, the singularity is globally naked, implying the existence of observers in the asymptotic region of the spacetime which ``see" the central singularity after a finite proper time. Although it is tempting to regard this fact as a violation of the Weak Cosmic Censorship (WCC) conjecture, a genuine violation would only occur if globally naked singularities persisted under generic perturbations of the TB model, taking into account realistic effects including non-spherical deformations of the initial data, angular momentum and pressure gradients. This has motivated researchers to investigate the behaviour of several classes of perturbations in the vicinity of the central singularity, including the propagation of electromagnetic radiation in the geometric optics approximation~\cite{dC84,bWkL89,iD98,nOoS14}, the propagation of test scalar fields on TB backgrounds~\cite{bNtW02,nOoS14}, and linearized gravitational perturbations with odd~\cite{hItHkN98,hItHkN99,eDbN11b} and even~\cite{hItHkN00,tWbN09,eDbN11} parity. (For studies regarding the behaviour of quantized fields on TB spacetimes, see for instance Refs.~\cite{sBtScVlW98,sBtScVlW98b,sBtScV00}.) Despite of these efforts, it is still not completely clear whether or not TB globally naked singularities survive under generic perturbations, and thus it constitutes a challenging open problem in theoretical physics. This problem is of particular interest since the TB spacetime is an example of a four dimensional, asymptotically flat spacetime which admits the dynamical formation of globally naked singularities\footnote{WCC violation has also been studied in five~\cite{lLfP10,yYhS11} and higher-dimensional spacetimes~\cite{rGpJ04,rGpJ07}. Remarkably, WCC violation has been recently discovered in a six-dimensional, asymptotically flat spacetime~\cite{pFmKlLsT17}.}.

With respect to the stability problem of the Cauchy horizon associated with globally naked singularities within the TB collapse model, in a previous paper~\cite{nOoS14} we have addressed the case of spherical test scalar fields. In particular, we have proven that such fields cannot grow arbitrarily large as they propagate outwards along the Cauchy horizon. This work constitutes an extension of our analysis, where we have studied the behaviour of test fields with arbitrary angular momentum as well as odd-parity linearized gravitational perturbations. Our most important result, Theorem~\ref{Thm:Main}, is presented in section~\ref{Sec:EnergyEstimates}, where we considered initially smooth test fields with associated zero-divergence stress-energy tensor satisfying the dominant energy condition, and where we proved that such fields possess a positive-definite energy norm which is uniformly bounded on the domain of dependence. In particular, our result implies that the field energy measured by free-falling observers co-moving with the dust particles is bounded as the observers cross the Cauchy horizon, even if the crossing point lies arbitrarily close to the central singularity. We would like to emphasize that the existence of a uniformly bounded energy norm for the field is remarkable given the fact that the background geometry is not only dynamical but singular. In section~\ref{Sub:Electromagnetic_fields}, our main result was applied to the propagation of electromagnetic fields, concluding that the total energy and radiation flux of the fields are bounded in the vicinity of the central singularity. Further, in section~\ref{Sub:Scalar_fields}, combining~Theorem~\ref{Thm:Main} with a Sobolev-type estimate, we obtained point-wise estimates which control the behaviour of the Klein-Gordon field close to the central singularity and the Cauchy horizon. Although these estimates leave open the possibility that the field diverges at the first singular point, they imply that the scalar field must decay to zero along the Cauchy horizon. Finally, in section~\ref{Sub:GravPert}, we considered odd-parity linear gravitational and dust perturbations of the collapsing spacetime. In this case, we proved that the relevant gauge-invariant combinations of the metric perturbations stay bounded everywhere on the domain of dependence with the possible exception of a small region close to the central singularity Our rigorous results are in accordance with previous studies in the marginally bound collapse case~\cite{hItHkN98,hItHkN99}, which showed numerical evidence that the perturbations are well behaved along the Cauchy horizon, even though the gravitational perturbation diverges at the central singularity. As mentioned above, our estimates do not exclude a divergence of the fields at the first singular point; nevertheless the numerical results in~\cite{hItHkN98,hItHkN99} and other numerical simulations performed by one of us, indicate that the divergence of the gravitational field close to the first singular point is slower than the one allowed by our bounds, indicating that they are not optimal and could be improved. Likely, obtaining improved bounds requires estimates on higher-oder derivatives of the fields. This will be explored in future work.

Further, it would be desirable to extend our analysis to linearized gravitational perturbations with even parity. In this case, the linearized equations are much more complicated than in the odd-parity sector since the perturbations of the dust fields are coupled to the metric ones, and hence it is a priori not clear if these equations possess an effective stress-energy tensor from which appropriate energy estimates can be constructed. To this respect, we note the prominent work by Duffy and Nolan~\cite{eDbN11} which provide a rigorous study for the stability of the Cauchy horizon under even-parity gravitational perturbations in the self-similar case, and concludes that gauge-invariant combinations of the metric perturbations diverge at the Cauchy horizon. A similar conclusion for even-parity quadrupolar gravitational perturbations in the marginally bound, non-self-similar case was obtained earlier by Iguchi, Harada and Nakao~\cite{hItHkN00} based on numerical work. For these reasons, it would be interesting to generalize our analysis to the even-parity sector.

Besides the analysis for the behaviour of test fields and linearized odd-parity gravitational perturbations analyzed in this article and our previous one~\cite{nOoS14}, it would also be interesting to explore whether or not the stability of a naked singularity might be related to its strength, as originally defined by Tipler~\cite{fT77} and further analyzed and generalized in Refs.~\cite{cCaK85,rN86,pSaL99,bN99,bN00,aO00}. As can be inferred from these references, the central naked singularity in the TB collapse model, under our assumptions on the initial data, is Tipler-weak with respect to radial null geodesics; however, it is Tipler-strong with respect to the central timelike geodesic and deformationally strong along non-radial null geodesics terminating at the central singularity (see Proposition 1 in Ref.~\cite{bN00}). Based on these properties, one might be able to construct wave packets in the geometric optics approximation which are compressed in one direction as they approach the singularity, and study their effect on the stability of the naked singularity and the associated Cauchy horizon.

\acknowledgments
It is our pleasure to thank Mihalis Dafermos, Brien Nolan, and Thomas Zannias for suggestions and stimulating discussions. We also thank an anonymous referee for pointing out to us the notion of Tipler-strength of a singularity, which may be relevant for future projects related to this article. This work was supported in part by CONACyT Grants No. 46521 and 232390, by the CONACyT Network Project 280908 ``Agujeros Negros y Ondas Gravitatorias", and by a CIC Grant to Universidad Michoacana. OS thanks the Perimeter Institute, where part of this work was completed, for hospitality. Research at the Perimeter Institute is supported by the Government of Canada through Industry Canada and by the Province of Ontario through the Ministry of Research and Innovation.

\appendix
\section{Proof of Lemma~\ref{Lem:Technical}}
\label{App:Technical}

The goal of this appendix is to establish the bound
\begin{equation}
\tau_s(R_{CH}(\tau)) - \tau \geq K(\tau - \tau_s^0)^{6/7},\qquad
\tau_s^0 < \tau < \tau_s^0 + \delta,
\end{equation}
for some positive constants $K > 0$ and $\delta > 0$, where here the function $R_{CH}: [\tau_s^0,\infty)\to\Real$ parametrizes the location of the Cauchy horizon $(\tau,R_{CH}(\tau))$ in the $(\tau,R)$-coordinates, and $\tau_s(R)$ defined in Eq.~(\ref{Eq:taus}) is the collapse time for the dust shell $R$.

For this, we work in the small region $D(\varepsilon)$ defined in Eq.~(\ref{Eq:DepsDef}) and start with a bound on $\tau_{CH}(R)$, where $\tau_{CH}(R)$ is the proper time at which the dust shell $R$ intersects the Cauchy horizon. Using Lemma~\ref{Lem:Bounds} and Eq.~(\ref{Eq:yCH}) we find
\begin{equation}
\tau_s(R) - \tau_{CH}(R) \geq \frac{2}{3}\frac{y_{CH}(R)^3}{\sqrt{c(R)}}
 = \frac{\Lambda_0}{2\sqrt{c_0}} R^2\left[ 1 
  - \frac{3\sqrt{c_0}}{2}\left( \frac{3\Lambda_0}{4} \right)^{-1/3} R^{1/3}
  + {\cal O}(R^{2/3}) \right],
\label{Eq:ABound1}
\end{equation}
where $c_0 := c(0) > 0$ and $\Lambda_0 = \Lambda(0,0)$. On the other hand, according to Eq.~(\ref{Eq:dtau}) we have
$$
\tau_s(R) = \tau_s^0 + \frac{\Lambda_0}{2\sqrt{c_0}} R^2 + {\cal O}(R^3),
$$
and thus
$$
\tau_{CH}(R) - \tau_s^0 \leq \left( \frac{3\Lambda_0}{4} \right)^{2/3} R^{7/3}
 + {\cal O}(R^{8/3}).
$$
Hence, choosing $\varepsilon > 0$ sufficiently small in the region $D(\varepsilon)$ there exists a constant $k > 0$ such that
$$
(\tau_{CH}(R) - \tau_s^0)^{3/7} \leq k R
$$
for all $0\leq R \leq R(\varepsilon)$. Substituting $R = R_{CH}(\tau)$ we find
$$
R_{CH}(\tau) \geq \frac{1}{k}(\tau - \tau_s^0)^{3/7}
$$
for all $\tau_s^0\leq \tau \leq \tau_s^0 + \delta$ with $\delta > 0$ small enough. Substituting $R = R_{CH}(\tau)$ into Eq.~(\ref{Eq:ABound1}) and using the properties of $D(\varepsilon)$ we find
$$
\tau_s(R_{CH}(\tau)) - \tau 
 \geq \frac{\Lambda_0}{2\sqrt{c_0}}(1 - \varepsilon) R_{CH}(\tau)^2
 \geq \frac{\Lambda_0}{2\sqrt{c_0}}\frac{1 - \varepsilon}{k^2}(\tau - \tau_s^0)^{6/7},
$$
for all $\tau_s^0\leq \tau \leq \tau_s^0 + \delta$, which establishes the desired bound.

\section{Proof of Lemma~\ref{Lem:Sobolev} (Sobolev estimate)}
\label{App:Sobolev}

Let $r_1\geq 0$, $\Sigma := (r_1,\infty) \times S^2$, and denote by $X$ the complex vector space consisting of $C^\infty$-differentiable functions $\Phi: \Sigma\to \Complex$ such that $\Phi(r,\hat{x}) = 0$ for all $(r,\hat{x})\in (r_1,\infty)\times S^2$ with large enough $r$. For $\sigma \geq 0$, let $D^{\sigma}: X\to X$ be the linear operator defined by
\begin{equation*}
D^{\sigma} \Phi := \sum_{\ell m}(2\ell +1)^{\sigma} \phi_{\ell m} Y^{\ell m}, \qquad \Phi = \sum_{\ell m} \phi_{\ell m}Y^{\ell m},
\end{equation*}
where $Y^{\ell m}$ denote the standard spherical harmonics. Fix $\ell m$ and let $v := \phi_{\ell m} \in \Complex$, then for all $r \geq r_1$,
\begin{equation*}
r|v(r)|^2 = -\int_r^{\infty} \frac{d}{ds} s |v(s)|^2 ds 
 = -\int_r^{\infty} \left[ |v(s)|^2 + 2s\re(\bar{v}(s) v'(s)) \right] ds,
\end{equation*}
where the prime denotes differentiation with respect  to $r$. Multiplying both sides of this equation by $\Lambda_{\ell}^{1+2\sigma}$ where $\Lambda_{\ell} := 2\ell + 1$, we obtain
\begin{eqnarray}
\Lambda_{\ell}^{1+2\sigma} r|v(r)|^2 
 &=& -\int_r^{\infty} \left[ \Lambda_{\ell}^{1+2\sigma} |v(s)|^2 
  + 2\re(\Lambda_{\ell}^{1+\sigma}\bar{v}(s) \Lambda_{\ell}^{\sigma}sv'(s)) \right] ds
\nonumber\\
&\leq& \int_{r_1}^\infty \left[ -|\Lambda_{\ell}^{1/2+\sigma} v(s)|^2 
 + |\Lambda_{\ell}^{1+\sigma}v(s)|^2 + s^2|\Lambda_{\ell}^{\sigma}v'(s)|^2 \right] ds,
\label{Eq:Sobolev1}
\end{eqnarray}
where in the second step we have used the inequality $2\re(\bar{a}b) \leq |a|^2 + |b|^2$, $a,b \in \Complex$. For $\ell=0$ we have $\Lambda_\ell = 1$ and this inequality reduces to
$$
r|v(r)|^2 \leq \int_{r_1}^\infty s^2 |v'(s)|^2 ds,
$$
which proves the lemma for the spherically symmetric part, $\Phi_{\ell=0}$, of the field. For the remaining part, we discard the negative term on the right-hand side of Eq.~(\ref{Eq:Sobolev1}) and sum over $\ell m$. Observing that
\begin{equation*}
\langle |D^{\sigma}\Phi|^2 \rangle_{S^2} 
 = \sum_{\ell m} | \Lambda_{\ell}^{\sigma} \phi_{\ell m} |^2,
\end{equation*}
we obtain
\begin{equation}
\langle r |D^{\kappa}\Phi|^2 \rangle_{S^2} 
 \leq \int_{r_1}^\infty \langle r^2|D^\sigma\Phi'|^2 + |D^{1+\sigma}\Phi|^2 \rangle_{S^2} dr,
\label{Eq:app_inequality_1}
\end{equation}
with $\kappa := 1/2 + \sigma > 1$.

To complete the proof of the Lemma, it remains to show that the left-hand side of Eq.~(\ref{Eq:app_inequality_1}) is an upper bound for $r|\Phi|^2$. In order to prove this, we use the fact that $4\pi \sum_{m=-\ell}^{\ell} |Y^{\ell m}|^2 = \Lambda _{\ell}$ and set $a_{\ell} := \left( \sum_{m=-\ell}^{\ell} |\phi_{\ell m}|^2 \right)^{1/2}$. Then, for each fixed $r \geq r_1$ and $\hat{x} \in S^{2}$, the Cauchy-Schwarz inequality leads to
\begin{equation*}
|\Phi|^2 \leq \left( \sum_{\ell m} |\phi_{\ell m}||Y^{\ell m}| \right)^2 \leq \frac{1}{4\pi}\left( \sum_{\ell=0}^{\infty} \Lambda_{\ell}^{1/2-\kappa} \Lambda_{\ell}^{\kappa}a_{\ell} \right)^2 \leq \frac{1}{4\pi}\left( \sum_{\ell=0}^{\infty}\Lambda_{\ell}^{1-2\kappa} \right)\left( \sum_{\ell=0}^{\infty}|\Lambda_{\ell}^{\kappa}a_{\ell}|^2 \right).
\end{equation*}
The sum inside the first parenthesis on the right-hand side converges because $\kappa > 1$. The expression inside the second parenthesis on the right-hand side is equal to $\langle |D^{\kappa} \Phi|^2 \rangle_{S^2}$. Therefore, we conclude that
\begin{equation}
|\Phi(r,\hat{x})|^2 \leq C_{\kappa} \langle |D^{\kappa}\Phi|^2 \rangle_{S^2}
\label{Eq:app_inequality_2}
\end{equation}
for all $(r,\hat{x}) \in \Sigma_{\tau}$, with $C_{\kappa} := \sum_{\ell=0}^{\infty}\Lambda_{\ell}^{1-2\kappa}/4\pi$. Now the statement of the lemma follows from the inequalities~(\ref{Eq:app_inequality_1})~and~(\ref{Eq:app_inequality_2}).
\qed

\bibliographystyle{unsrt}
\bibliography{refs_collapse}

\end{document}